\documentclass[aps,prx,twocolumn,amsmath,nofootinbib,10pt,floatfix]{revtex4-2}

\usepackage[english]{babel}
\usepackage{graphicx}
\usepackage{ulem}
\usepackage{soul}
\usepackage{float}
\usepackage[colorlinks, linkcolor=red , citecolor= blue,breaklinks=true]{hyperref}
\usepackage{bm}
\usepackage{bbm}
\usepackage{verbatim}
\usepackage{esint}
\usepackage{marginnote}
\usepackage{feynmf}
\usepackage{esint}

\newcommand{\RNum}[1]{\uppercase\expandafter{\romannumeral #1\relax}}

\def \beq {\begin{eqnarray}}
\def \eeq {\end{eqnarray}}

\begin{document}
\title{Universal Scaling Solution for a Rigidity Transition: Renormalization Group Flows Near the Upper Critical Dimension}
\author{Stephen J. Thornton}
\affiliation{Department of Physics, Cornell University, Ithaca NY 14853, U.S.A.}
\author{Danilo B. Liarte}
\affiliation{Institute of Theoretical Physics, S\~ao Paulo State University, S\~ao Paulo, SP, Brazil}
\affiliation{ICTP South American Institute for Fundamental Research, S\~ao Paulo, SP, Brazil}
\affiliation{Department of Physics, Cornell University, Ithaca, NY 14853, USA}
\author{Itai Cohen}
\affiliation{Department of Physics, Cornell University, Ithaca NY 14853, U.S.A.}
\author{James P. Sethna}
\affiliation{Department of Physics, Cornell University, Ithaca NY 14853, U.S.A.}
\date\today
\begin{abstract}
Rigidity transitions induced by the formation of system-spanning disordered rigid clusters, like the jamming transition, can be well-described in most physically relevant dimensions by mean-field theories. A dynamical mean-field theory commonly used to study these transitions, the coherent potential approximation (CPA), shows logarithmic corrections in $2$ dimensions. By solving the theory in arbitrary dimensions and extracting the universal scaling predictions, we show that these logarithmic corrections are a symptom of an upper critical dimension $d_\mathrm{upper}=2$, below which the critical exponents are modified. We recapitulate Ken Wilson's phenomenology of the $(4-\epsilon)$-dimensional Ising model, but with the upper critical dimension reduced to $2$. We interpret this using normal form theory as a transcritical bifurcation in the RG flows and extract the universal nonlinear coefficients to make explicit predictions for the behavior near $2$ dimensions. This bifurcation is driven by a variable that is dangerously irrelevant in all dimensions $d>2$ which incorporates the physics of long-wavelength phonons and low-frequency elastic dissipation. We derive universal scaling functions from the CPA sufficient to predict all linear response in randomly diluted isotropic elastic systems in all dimensions.
\end{abstract}
\maketitle

\section{Introduction}\label{Intro}

We present here a complete analysis of a particular
isotropic, homogeneous
rigidity transition. Our solution provides universal scaling predictions for
the linear responses of the system -- viscosities, elastoplastic and
viscoelastic functions, Green's functions, densities of states, etc. It also
implies a renormalization-group flow which recapitulates the classic 
$(4-\epsilon)$-dimensional Ising critical point predictions with the upper
critical dimension reduced to two. 

There is a family of rigidity transitions of past and current interest, 
with many common features but not all sharing the same universality class.
Most prominant is the recent focus on the jamming
transition~\cite{LiuNag2010,BehringerCha2018}, applied to glasses,
colloidal and granular systems, and foams. In jamming, a replica solution
in infinite dimensions~\cite{charbonneau2017glass,parisi2020theory} makes quantitative predictions
for microscopic power laws (universal contact force and gap size distributions)
all the way down to two dimensions~\cite{Charbonneau2012UniversalMA,
Charbonneau2015JammingCR,sartor2021mean}. Other examples include
rigidity transitions in tissues~\cite{manning16,heisenberg21} relevant to
wound healing and embryonic organ formation, dislocation entanglement in
crystals~\cite{MiguelZap2002}, and `double descent' accuracy transitions in
deep learning~\cite{BahriGan2020}. Several of these systems appear to share the same dimension-independent universal power laws above two dimensions, motivating the simplified model investigated here. 

Our work is inspired by the discovery~\cite{LiarteXOL19} of diluted networks in both two and three dimensions that show a jamming transition as an endpoint of a line of rigidity percolation transitions, studied through static simulations and via the coherent potential approximation. These networks have no linear elastic moduli in a floppy phase, and show a jump in the bulk modulus and linear growth in the shear modulus at a jamming point of the phase boundary. With the inclusion of additional angular and bending forces, these models are also believed to be applicable to fiber networks, such as the ones found in cytoskeletal networks and extracellular matrices~\cite{CytoWeitz2004,FredCyto2003}. Here, we present an analysis of this model for the rigidity percolation transition, where all static moduli grow linearly from zero.

Our model is an isotropic, continuum version of the coherent potential approximation (CPA)~\cite{FengGar1985,LiarteXOL19,LiarteTSCCS22}. It replicates the CPA predictions for the dilution of a random, amorphous spring network published by D\"uring et al.~\cite{During2013}. Both are outgrowths of what is termed rigidity percolation (RP)~\cite{thorpe1985rigidity,thorpe1999rigidity,jacobs1995generic}, where a network of springs connecting nodes with no angular forces is diluted, not until it becomes completely disconnected (percolation) but until its elastic moduli vanish. It is known that rigidity percolation on a two-dimensional triangular lattice has critical exponents that differ from those found for rigidity percolation on graphs generated from jammed packings of spheres, with a modulus that with a higher power with excess coordination number~\cite{BroederszMac2011, WillWIP}. We conjecture that lattice anisotropy~\cite{DiffusionLimitedAggregation}, and/or undeformed springs in a line (forming \textit{second-order constraints}~\cite{jacobs1995generic,ManningSantangelo}) are relevant perturbations at the rigidity transition for spring networks, and that our theory is applicable to a randomly diluted isotropic network without these buckling transitions (Section~\ref{Applicability}). This continuum theory, when used to describe jamming~\cite{LiarteXOL19,ThorntonLSWIP}, matches most of the properties seen numerically, both directly and in spring network models generated from jammed configurations~\cite{GoodrichD-BSTvHLN14} and simulations of diluted amorphous spring networks~\cite{During2013}. The calculation presented here focuses on the case where the bulk modulus does not jump, which includes most of the rigidity transitions other than jamming. Nevertheless, this calculation captures the predicted scaling of the shear mode close to the jamming transition~\cite{ThorntonLSWIP}. We leverage our exact solutions of the continuum theory to generate universal scaling forms for all linear response properties, and we use them to calculate new critical exponents below $2$ dimensions described by nonlinear renormalization group flows. We identify a scaling variable which is dangerously irrelevant for $d>2$ which is responsible for low-frequency dissipation, the phonon density of states, and the logarithmic corrections in $d=2$.

We conjecture that many qualitative features of our analysis are
important predictions and tools that should apply more generally. 
(1)~Many replica-theory and other mean-field methods yield self-consistent 
formulas that predict power-law dependence near
transitions~\cite{KurchanPrivateCommunication}. Our analysis is a guide
to extracting universal predictions out of these self-consistent relations. For example, the Curie-Weiss law predicts the entire phase behavior of 
an Ising magnet above four dimensions~\cite{SethnaBrazilLecture1},
but only the power laws and the scaling function
near the critical point are expected to be universal. 
(2)~Much impressive work, especially in jamming, has focused on the 
microscopic behavior~\cite{manning2015random,Wyart2012MarginalSC,Lerner2013LowenergyNE,Combe2000StrainVS,Jin2021AJP,cates1998jamming,o2001force}
(how jammed systems are \textit{different} from regular
systems~\cite{PWAndersonLesHouches78}). Our previous
work~\cite{GoodrichLS16,BaityJesiGLNS17,LiarteTSCCS22,LiarteTSCCSnn2,TLASC23} has taken a different perspective by analyzing the emergent phase behavior in terms of Widom scaling theories
(investigating how jamming may be \textit{reduced} to regular system behavior).
Our work here builds upon these by presenting a wonderful example of the rich phenomena that
can be extracted by focusing on the macroscopic behaviors in space and time of this and other systems.
(3)~Our extraction of
RG flows allows us to explain the quite non-trivial invariant scaling variables in the 
upper critical dimension (as we found also in the 4D Ising
model~\cite{ArchishmanNF}); these variables should appear in a wide class of models that share this upper critical dimension. 
(4)~Many higher-temperature features of glasses are missed in the
infinite-dimensional replica theories (such as the continued relaxation
below the mode-coupling transition) and are thought to be non-perturbative in the inverse dimension $1/d$. Our rigidity transition
has a dangerously irrelevant variable above two dimensions that is needed
to incorporate low-frequency vibrational modes and dissipation, which we show is
indeed such a non-perturbative effect in the limit $d\to\infty$. The work presented here provides a road map for dealing with such irrelevant variables.

Our work is similar in some aspects to that of a preprint by Vogel et al.~\cite{Fuchs2024arxiv}. They form a self-consistent theory for the shear response of a nearly unjammed solid with similar structure to our continuum CPA. Their analysis incorporates nontrivial momentum dependence which is important in the disordered glassy phase studied by the authors of the article in previous work~\cite{MaierZF18,ZippeliusSethnaTextbook2023} and by others~\cite{NampoothiriWRZBC20}; this type of momentum-dependent modulus is ignored by the CPA. Our work, on the other hand, is focused on calculating the universal scaling functions for the transition, deducing nontrivial normal forms for renormalization group flows, and understanding specifically the singular behavior of the theory found in two dimensions.

The organizational structure of our paper is as follows: in Section~\ref{CPA}, we briefly review the CPA as applied to weakened elastic media. We show that, under quite general assumptions, elastic moduli vanish linearly in deviations from the critical dilution fraction $\mu\sim\delta p\sim\delta z$. In Section~\ref{23Dim}, we evaluate the universal scaling functions for the space-time linear response of the theory near the critical point directly in $d=3$ dimensions and in $d=2$ dimensions, and show that the appropriate scaling for the dynamical behavior close to the critical point has log corrections in two dimensions, as seen earlier~\cite{During2013}. In Section~\ref{GenDim} and Appendix~\ref{AppC}, we cast solutions close to the critical point into a scaling theory in general dimensions $d$, and show that $d=2$ demarcates the boundary between two differing sets of critical exponents. We construct renormalization group flow equations that are consistent with the critical exponents predicted by the theory (with more details located in Appendices~\ref{AppA} and~\ref{AppB}), with an exchange of stability between two RG fixed points in $d=2$. In this way, we interpret $d=2$ as the upper critical dimension of the transition, and find the appropriate scaling variables in $d=2$ analogously to those in the 4D Ising model. The logarithmic shifts that are one signature of the upper critical dimension are investigated in more detail in Appendix~\ref{AppD} through a direct comparison to a lattice CPA calculation on the bond-diluted triangular lattice. Finally, in Section~\ref{Applicability}, we discuss the applicability of this model more broadly to a wide variety of different rigidity transitions in disordered elastic systems.

\section{The CPA and critical exponents for static moduli}\label{CPA}

\begin{figure}[ht]
\begin{center}
\includegraphics[width=\linewidth]
{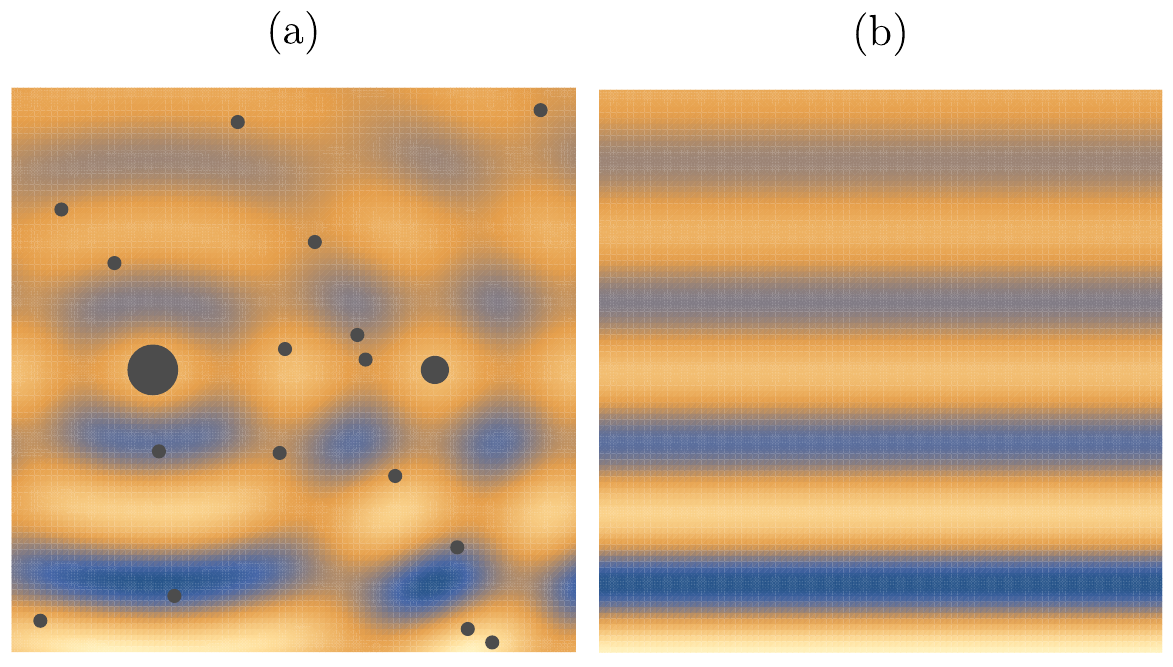}
\end{center}
\caption{\textbf{Schematic of the approximation made by the CPA}. (a) A phonon travelling (upwards) into an elastic medium with randomly distributed defects scatters in complicated ways. The dark circles represent the isotropically scattering  defects in our continuum CPA, while the colors represent the magnitude of the displacement field. (b) The same phonon travelling through the effective medium is damped as it propagates. The CPA gives renormalized elastic constants for a medium with no defects, which incorporates the strong scattering of shorter-wavelength phonons through a damping term in the effective moduli.}
\label{CPASchematic}
\end{figure}

We examine a continuum version of the coherent potential approximation (CPA) inspired by the lattice CPA~\cite{FengGar1985}. 
The lattice CPA can be used to describe a system comprised of purely harmonic springs of strength $k_0$ on some regular lattice that are independently randomly occupied with probability $p$. This is equivalent to placing a probability distribution on the strengths of bonds $k'$
\begin{equation}
k'\sim p\;\delta(k'-k_0)+\left(1-p\right)\;\delta\left(k'\right)
\end{equation}
so that each bond of strength $k_0$ is independently randomly occupied with probability $p$. Other effective medium theories have placed more realistic distributions of bonds based on observations of stress and strain fluctuations from simulations of particular systems (soft spheres touching, gels near their gelation point, etc.~\cite{EMTRandRegular}). One then tries to describe this disordered elastic system by a non-disordered \textit{effective medium}, whose physical properties are renormalized by $p$. Finding the best effective value of stiffness $k$ so that the disorder-averaged elastic Green's function $\left\langle \textbf{G}(k',\omega)\right\rangle\approx \textbf{G}\left(k\left(\omega\right),\omega\right)$, the elastic Green's function for an effective medium with no disorder, amounts to solving a self-consistent equation for the stiffnesses. The effective moduli are allowed to become frequency-dependent and complex, transforming them into \textit{visco}elastic moduli \cite{FengGar1985}. This CPA is related to the CPA used in other impurity scattering problems which makes a similar assumption that the self-energy is local $\Sigma\left(\omega,q\right)\approx\Sigma(\omega)$. The content of the approximation made by the CPA is twofold: first, the effective stiffnesses depend only on $\omega$, and not on $q$. Second, this constraint is imposed by requiring that the \textit{single-site} $T$-matrix for multiple scattering vanishes~\cite{KakehashiPRB2002}, as opposed to the \textit{full} $T$-matrix (which is analytically intractable). The result of the approximation is a homogeneous effective medium that incorporates the effects of phonons scattering off of defects introduced by the disorder into a effective damping (Figure~\ref{CPASchematic}).

The self-consistent equation for the shear modulus of the effective medium under the assumptions of the lattice CPA \cite{FengGar1985,LiarteXOL19} is
\begin{equation}\label{eq:SC}
\frac{p-\mu/\mu_{\textsc{F}}}{1-\mu/\mu_{\textsc{F}}}=\frac{1}{\widetilde{z}}\fint_{\textrm{BZ}}\textrm{d}^dq\;\textrm{Tr}\left(\textbf{D}\textbf{G}\right),
\end{equation}
where the integral is an average over the first Brillouin zone, $\widetilde{z}$ denotes the average number of constraints per microscopic unit in the undiluted system (in the lattice case, it is the number of bonds per node), $\mu_{\textsc{F}}$ is the shear modulus in the completely filled, non-disordered system, and $\textbf{D}$ and $\textbf{G}$ are the dynamical matrix and Green's function for the medium, respectively. This long-wavelength limit of the lattice problem reproduces the isotropic CPA for amorphous spring networks investigated by D\"uring et al. in~\cite{During2013}. This expression is self-consistently solved for $\mu\left(p,\omega\right)$, which is proportional to the best effective value of the microscopic stiffness $k$.

We are interested in properties of disordered rigidity transitions for systems that are statistically isotropic. To lose reference to any particular lattice, we pass to a continuum version of the CPA, where the Brillouin zone is replaced by a sphere of radius $q_{\textsc{D}}$ (the Debye wavevector) and dynamical matrices and Green's functions are written for an isotropic continuum elastic sheet. We will see that this version of the CPA also describes the CPA scaling behavior of diluted lattices with a continuous rigidity transition that are isotropic at long wavelengths, as anisotropic terms enter as corrections to scaling. With only one independent microscopic stiffness, it suffices to track the behavior of the renormalized shear modulus $\mu$ as it deviates from its value in the unweakened system $\mu_{\textsc{F}}$, as all other stiffnesses are proportional to this modulus. We decompose the continuum dynamical matrix and Green's function into transverse and longitudinal parts to evaluate the integrand:
\begin{equation}
\begin{split}
\textbf{D}&= D_{\textrm{L}}(q)\hat{q}_i\hat{q}_j+D_{\textsc{T}}(q)\left(\delta_{ij}-\hat{q}_i\hat{q}_j\right), \\
\textbf{G}&= G_{\textrm{L}}(q,\omega)\hat{q}_i\hat{q}_j+G_{\textsc{T}}(q,\omega)\left(\delta_{ij}-\hat{q}_i\hat{q}_j\right), \\
\textrm{Tr}\left(\textbf{D}\textbf{G}\right)&=G_{\textsc{L}}\left(q,\omega\right)D_{\textsc{L}}\left(q\right)+\left(d-1\right)G_{\textsc{T}}\left(q,\omega\right)D_{\textsc{T}}\left(q\right).
\end{split}
\end{equation}
At zero frequency, assuming the dynamical matrix of the effective medium is invertible (as it is on the solid side of the transition), $\textbf{G}(q,0)=\textbf{D}^{-1}$ (see the beginning of Appendix~\ref{AppC} for details on the specific forms of $D_{\textsc{L}/\textsc{T}}$ and $G_{\textsc{L}/\textsc{T}}$), and so the integrand is the trace of a $d$-dimensional identity matrix and the self-consistent equation can be evaluated directly:
\begin{equation}
\frac{p-\mu/\mu_{\textsc{F}}}{1-\mu/\mu_{\textsc{F}}}=\frac{d}{\widetilde{z}}\implies\frac{\left(p-d/\widetilde{z}\right)-\left(1-d/\widetilde{z}\right)\mu/\mu_{\textsc{F}}}{1-\mu/\mu_{\textsc{F}}}=0.
\end{equation}
The constant $d/\widetilde{z}$ is identified as $p_c$. It is identical to the Maxwell counting constraint ignoring states of self-stress: each microscopic unit has on average $p\widetilde{z}$ constraints and $d$ degrees of freedom, and so $p_c=d/\widetilde{z}$. Keeping $0<p_c<1$, and defining $\mu'=\left(1-d/\widetilde{z}\right)\mu/\mu_{\textsc{F}}$, one has
\begin{equation}
\frac{(p-p_c)-\mu'}{1-\mu'/(1-p_c)}=0,
\end{equation}
and so $\mu'\sim\left(p-p_c\right)^{f_{\textrm{CPA}}}$ with $f_{\textrm{CPA}}=1$: the static shear modulus vanishes linearly with $\delta p$ upon approaching the transition.

It is tempting, then, to declare the CPA a mean-field-like theory that gives dimension-independent critical exponents, as Landau theory does in magnets. This turns out not to be true; the dynamical scaling of the theory has much more interesting structure.
\section{Two and three dimensions and a dangerously irrelevant variable}\label{23Dim}
In this section, we expand the solution for the viscoelastic modulus $\mu\left(\omega\right)$ close to its continuous stiffness transition. We then cast the solution near the critical point in terms of scaling variables and identify universal scaling functions. In three dimensions (Section~\ref{subsec:3d}), we show that it is necessary to retain an invariant scaling combination associated with an irrelevant variable to capture the low-frequency dissipative part of the viscoelastic modulus in the case of microscopically undamped dynamics. This irrelevant variable also gives information about the low-frequency density of states. This suggests that the irrelevant variable is \textit{dangerous}: it vanishes under a coarse-graining procedure, but it cannot be set to $0$ directly without losing access to a description of important low-frequency vibrational modes.

In two dimensions (Section~\ref{subsec:2d}), we show that the scaling variables that were correct in three dimensions no longer capture the behavior near the critical point. There are large, logarithmic shifts in physically relevant frequencies. In Section~\ref{GenDim}, we will show that this modification of the scaling variables is a result of the leading irrelevant variable becoming marginal in $d=2$, significantly altering the low-energy physics. This identifies $d=2$ as the upper critical dimension, and we write new critical exponents and scaling functions below the upper critical dimension. We also construct renormalization group flows consistent with the analytic structure of the scaling variables.

\subsection{Scaling in 3 dimensions}
\label{subsec:3d}
We first note that, at zero frequency, $\mu$ vanishes linearly with $p$ as $p\rightarrow 3/\widetilde{z}\equiv p_c$. We subtract $p_c$ from each side of Equation~\eqref{eq:SC} so that the self-consistent relation becomes
\begin{equation}
\begin{split}
\frac{(p-p_c)-\left(1-p_c\right)\mu/\mu_{\textsc{F}}}{1-\mu/\mu_{\textsc{F}}}&=\\
\frac{3}{\widetilde{z}q_{\textsc{D}}^3}\left(\int_0^{q_{\textsc{D}}}\textrm{d}q\frac{wq^2}{\left(\lambda_{\textsc{F}}/\mu_{\textsc{F}}+2\right)\mu q^2-w}\right.&\left.+2\int_0^{q_{\textsc{D}}}\textrm{d}q\frac{wq^2}{\mu q^2-w}\right)
\end{split}
\end{equation}
where $w$ is set by the dynamics. We focus on undamped dynamics, where $w=\rho\omega^2$, but $w=i\gamma\omega$ for overdamped dynamics has very similar scaling behavior. One could in principle also consider the case of Galilean-invariant Kelvin damping, where $w=\rho\omega^2+i\eta\omega q^2$, but the analysis of the asymptotic scaling in this manuscript assumes $w$ is $q$-independent. We write a scaling theory for small $\delta p\equiv p-p_c$, which is the distance to the critical point as measured in $p$. At $w=0$, the integral terms vanish. This suggests a zero-frequency scaling for the shear modulus:
\begin{equation}\label{eq:Mscaling}
M\equiv \frac{\mu}{\mu_{\textsc{F}}/(1-p_c)}\frac{1}{\left|\delta p\right|}.
\end{equation}
The integrals can be done directly; it is useful to substitute $\xi=\left(q/q_{\textsc{D}}\right)^2$ and rescale to $w_{\textsc{L}}=w/\left(\lambda_{\textsc{F}}/\mu_{\textsc{F}}+2\right)q_{\textsc{D}}^2$ and $w_{\textsc{T}}\equiv w/q_{\textsc{D}}^2$ to find
\begin{equation}
\begin{split}
\frac{\delta p-\left|\delta p\right|M}{1-\left|\delta p\right|M/\left(1-p_c\right)}&=\\
-\frac{3}{2\widetilde{z}}\left(\int_0^{1}\textrm{d}\xi\frac{\xi^{1/2}}{1-\left(\mu/w_{\textsc{L}}\right)\xi}\right.&\left.+2\int_0^{1}\textrm{d}\xi\frac{\xi^{1/2}}{1-\left(\mu/w_{\textsc{T}}\right)\xi}\right).
\end{split}
\end{equation}
Assuming $\textrm{Im}\left(\mu\right)<0$ for $w>0$ (necessary for causality), we have
\begin{equation}
\begin{split}
\frac{\delta p-\left|\delta p\right|M}{1-\left|\delta p\right|M/\left(1-p_c\right)}&=\frac{3}{\widetilde{z}}\left(\frac{w_{\textsc{L}}}{\mu}+2\frac{w_{\textsc{T}}}{\mu}\right)-\\
-\frac{3}{\widetilde{z}}\left(\left(\frac{w_{\textsc{L}}}{\mu}\right)^{3/2}\tanh^{-1}\left(\sqrt{\frac{\mu}{w_{\textsc{L}}}}\right)\right.&+\\
+2\left(\frac{w_{\textsc{T}}}{\mu}\right)^{3/2}&\left.\tanh^{-1}\left(\sqrt{\frac{\mu}{w_{\textsc{T}}}}\right)\right).
\end{split}
\end{equation}
We are interested in the low-frequency behavior. The terms proportional to $w/\mu$ dominate over the terms proportional to $\left(w/\mu\right)^{3/2}$ at low frequencies. This suggests a scaling $w/\mu\sim\left|\delta p\right|$, which suggests we should expand the functions $\tanh^{-1}\left(z\right)$ about their appropriate complex infinity. For undamped dynamics, we note that $\textrm{Im}\left(\mu\right)\leq 0$ and hence $\textrm{Im}\left(\sqrt{\mu}\right)\leq 0$ to respect causality. The function $\tanh^{-1}(z)$ can then be expanded to find
\begin{equation}
\begin{split}
\frac{\delta p-\left|\delta p\right|M}{1-\left|\delta p\right|M/\left(1-p_c\right)}&\approx\frac{3}{\widetilde{z}}\left(\frac{w_{\textsc{L}}}{\mu}+2\frac{w_{\textsc{T}}}{\mu}\right)+\\
&+\frac{3}{\widetilde{z}}\frac{i\pi}{2}\left(\left(\frac{w_{\textsc{L}}}{\mu}\right)^{3/2}+2\left(\frac{w_{\textsc{T}}}{\mu}\right)^{3/2}\right).
\end{split}
\end{equation}
One can check that since $\textrm{Im}\left(\mu\right)\leq0$, we can write
\begin{equation}
\begin{split}
\frac{\delta p-\left|\delta p\right|M}{1-\left|\delta p\right|M/\left(1-p_c\right)}&\approx\frac{3}{\widetilde{z}}\left(\frac{w_{\textsc{L}}}{\mu}+2\frac{w_{\textsc{T}}}{\mu}\right)+\\
+\frac{3\pi}{2\widetilde{z}}&\left(\left(-\frac{w_{\textsc{L}}}{\mu}\right)^{3/2}+2\left(-\frac{w_{\textsc{T}}}{\mu}\right)^{3/2}\right).
\end{split}
\end{equation}
This expansion is more general and also works for the overdamped case where $w\sim i\gamma\omega$. This asymptotic analysis of the leading correction to the CPA result reproduces the calculation done in Appendix B of~\cite{DeGiuli2014}. Now we are prepared to define more invariant scaling variables and our first universal scaling function. We first insert the definition of $M$ in for $\mu$ everywhere. Then, we define $\Omega^2$ (named for the undamped case) as the scaling variable for $w$ which makes the dominant term of the scaling for the frequency-dependent part $\left|\delta p\right|\Omega^2/M$. The definition of $w$ in terms of $\Omega$ is then inserted in the higher-order frequency piece, and the remaining terms are all absorbed into a scaling combination $G$ for an irrelevant variable. The result, after dividing both sides by $\left|\delta p\right|$ and throwing away the higher-order contribution from the denominator of the left-hand side, is
\begin{equation}
\begin{split}
\pm1-M=\frac{\Omega^2}{M}+G\left(-\frac{\Omega^2}{M}\right)^{3/2},\\
\Omega^2\equiv\frac{\omega^2/\omega_0^2}{\left|\delta p\right|^2},\;\;\; G\equiv g/g_0\left|\delta p\right|^{1/2}.
\end{split}
\end{equation}
The sign $\pm1=\delta p/\left|\delta p\right|$ is for the rigid and floppy side of the transition, respectively; formulas for $\omega_0$ and $g/g_0$ can be found in Appendix~\ref{AppC}. This is an implicit definition of a universal scaling function for $M=\mathcal{M}\left(\Omega,G\right)$.%
\footnote{In a more complete self-consistent theory beyond the CPA such as~\cite{Fuchs2024arxiv}, the shear modulus will also have a $q$-dependence, leading to a scaling form $\mathcal{M}\left(\Omega,Q,G\right)$.} 
Setting the leading irrelevant piece $G$ to $0$ allows us to solve a quadratic equation for $M$ as $2\mathcal{M}\left(\Omega,0\right)=\pm1-\sqrt{1-4\Omega^2}$, as found indirectly in \cite{LiarteXOL19} and directly in \cite{LiarteTSCCS22,LiarteTSCCSnn2}. However, this form of the universal scaling function (unphysically) has no scattering-induced dissipation on the rigid side of the transition until $\Omega=1/2$, while the full solution retaining $G$ has $\textrm{Im}\left(\mathcal{M}\right)<0$ for all $\Omega>0$.%
\footnote{The effective medium theory misses the important contributions of quasilocalized modes to the low-frequency density of states, which give a characteristic scaling $D\left(\omega\right)\sim\omega^4$~\cite{Quasilocalized2018}.}
As mentioned in the introduction to this section, this identifies $G$ as an invariant scaling combination associated with a \textit{dangerously} irrelevant variable for the case of undamped microscopic dynamics, as the function $\mathcal{M}(\Omega,G)$ is not analytic in its second argument at zero. Retaining $G$ is necessary to understand the details of the low-frequency viscosity, susceptibility, and density of states. 

\begin{figure}[ht]
\begin{center}
\includegraphics[width=\linewidth]
{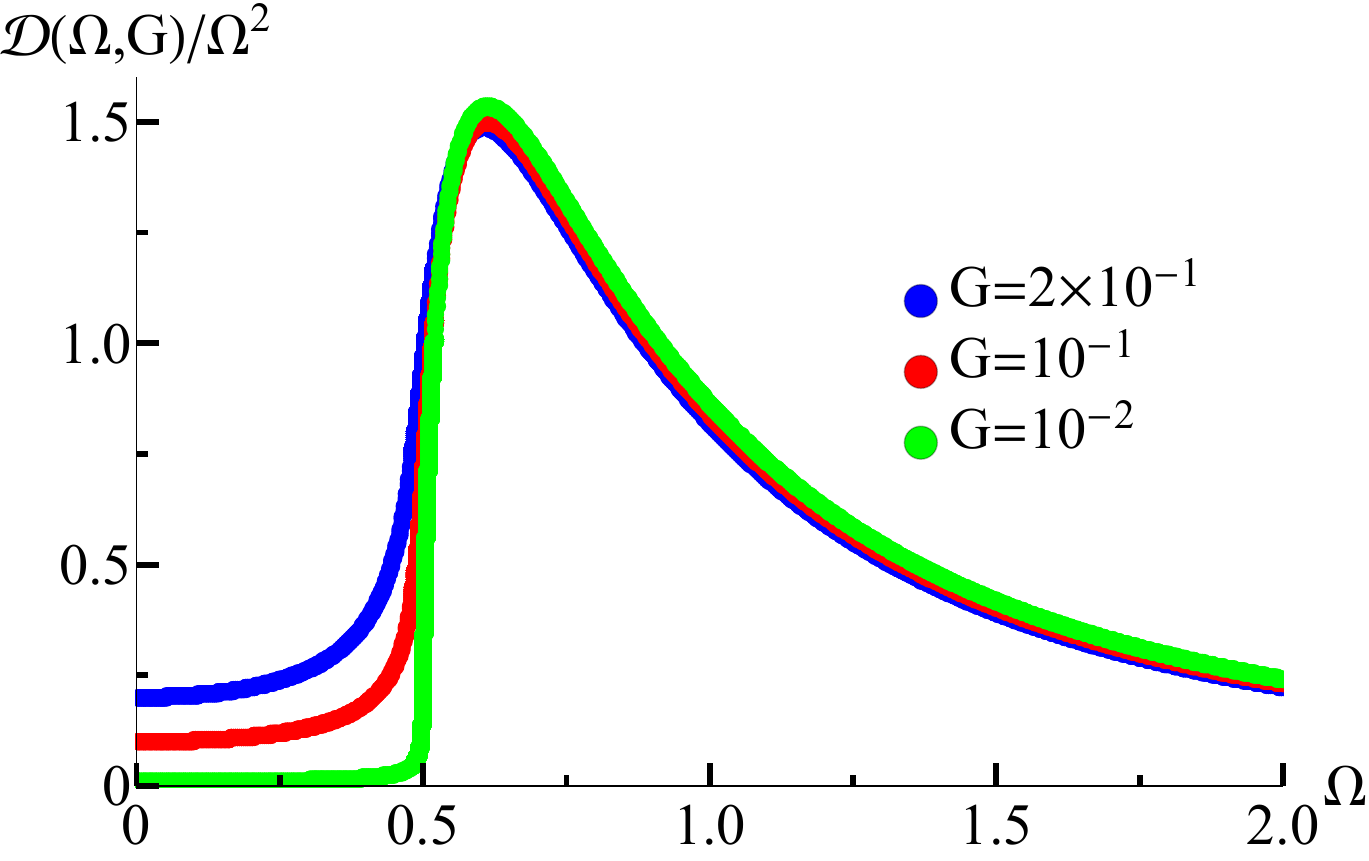}
\end{center}
\caption{\textbf{Prediction for the shape of the excess density of states near the rigid-floppy transition in $d=3$} (see also Fig.~\ref{DOS3DG}). We compare our density of states to the Debye result and find an excess of states that contribute to the \textit{boson peak} that is often seen in disordered rigid systems. The dangerously irrelevant variable $G$ must be retained to capture the Debye phonon contribution
to the density of states
below $\Omega=1/2$. These long-wavelength phonons are of course important
to the physics, but are swamped near the rigidity transition by the flat
density of states above $\omega^*$. Hence the phonon contribution is irrelevant in the RG sense, even though it is important to the physics.}

\label{BosonPeak}
\end{figure}

For models with many soft modes, we can illustrate the excess number of soft modes by comparing the density of states with that of the Debye model for a crystal, in which $D\left(\omega\right)\sim\omega^{d-1}$. We plot the predicted universal scaling forms for the density of states divided by the Debye form for different values of the dangerously irrelevant variable in Figure~\ref{BosonPeak}. The peak in the excess density of states is located near the frequency $\omega^*\sim\left|\delta p\right|$ where the density of states becomes nearly flat: a characteristic feature of all rigidity transitions in this family. The dangerous
irrelevant variable $G$ controls the contiuum phonon density of states, which
vanishes in the appropriate scaling limit.

We note in passing that the frequency scaling variable only appears in conjunction with the scaling variable for the modulus. Instead of defining a scaling for $w$, then, we could define a scaling variable for $f\equiv w/\mu$. This turns out to be a particularly natural choice that eases the analysis in two dimensions. Written in terms of this scaling variable, the self-consistent equation reads
\begin{equation}\label{eq:scaling3D}
\begin{split}
\pm1-M&=F+G\left(-F\right)^{3/2},\\
F&\equiv \frac{f/f_0}{\left|\delta p\right|}.
\end{split}
\end{equation}
\subsection{Scaling in 2 dimensions}
\label{subsec:2d}
\begin{figure}[ht]
\begin{center}
\includegraphics[width=\linewidth]
{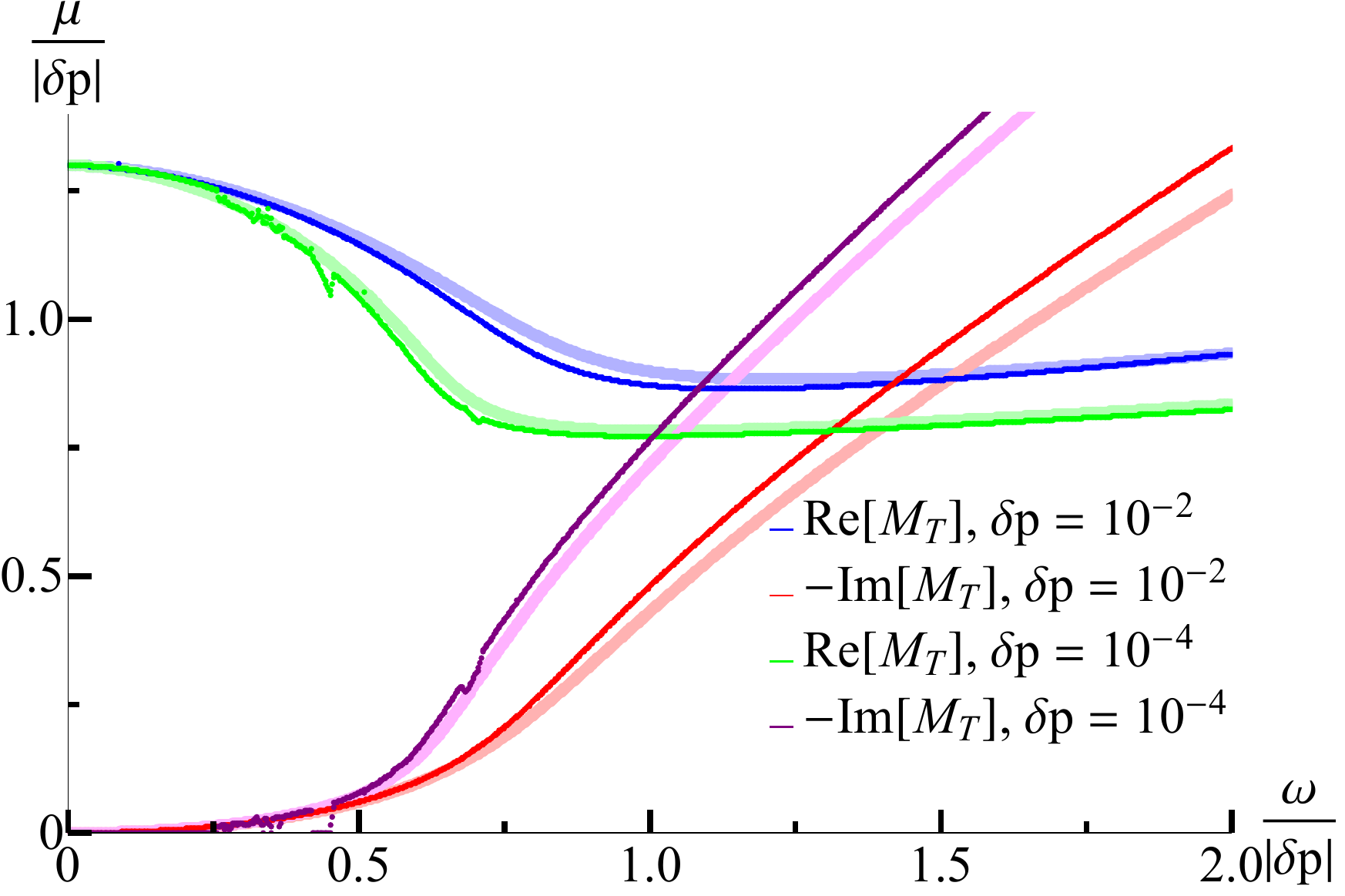}
\end{center}
\caption{\textbf{Logarithmic frequency shifts in an effective medium theory for 2D rigidity percolation on the triangular lattice}. We plot numerical solutions of the lattice CPA (thin lines), rescaled shear modulus $M_{\textsc{T}}\equiv\mu/\left|\delta p\right|$ as a function of rescaled frequency $\Omega_{\textsc{T}}\equiv\omega/\left|\delta p\right|$, against our scaling solutions (thick lines, Equations~\eqref{eq:2DUniversalSFM} and~\eqref{eq:2DTriangularSCM}). This comparison is done at two distances from the critical point $\delta p=\left\{10^{-2},10^{-4}\right\}$, demonstrating nice agreement. All parameters in our scaling form are determined from the long-wavelength parameters of the triangular lattice (Appendix~\ref{AppD}), so there are no fitting parameters. The shift in the rescaled frequency $\Omega_{\textsc{T}}^*$ where $\textrm{Re}\left(M_{\textsc{T}}\right)=-\textrm{Im}\left(M_{\textsc{T}}\right)$  from $\sim1.4$ to $\sim1.0$ upon reducing $\delta p$ from $10^{-2}$ to $10^{-4}$ is due to logarithmic corrections present in the upper critical dimension.}
\label{TriangularCPANumerics}
\end{figure}

We follow the steps above, and again evaluate the integral directly in $d=2$, where now $2/\widetilde{z}=p_c$. We subtract $p_c$ from each side of Equation~\eqref{eq:SC} so that the self-consistent relation becomes
\begin{equation}
\begin{split}
\frac{(p-p_c)-\left(1-p_c\right)\mu/\mu_{\textsc{F}}}{1-\mu/\mu_{\textsc{F}}}&=\\
\frac{2}{\widetilde{z}q_{\textsc{D}}^2}\left(\int_0^{q_{\textsc{D}}}\textrm{d}q\frac{wq}{\left(\lambda_{\textsc{F}}/\mu_{\textsc{F}}+2\right)\mu q^2-w}\right.&\left.+\int_0^{q_{\textsc{D}}}\textrm{d}q\frac{wq}{\mu q^2-w}\right)
\end{split}
\end{equation}
We again write a scaling theory for small $\delta p$. The zero-frequency scaling for the modulus is the same as in Eq. \eqref{eq:Mscaling}. We perform the same substitution as in $d=3$ of $\xi=\left(q/q_{\textsc{D}}\right)^2$ and rescale to $w_{\textsc{L}}=w/\left(\lambda_{\textsc{F}}/\mu_{\textsc{F}}+2\right)q_{\textsc{D}}^2$ and $w_{\textsc{T}}\equiv w/q_{\textsc{D}}^2$ to find
\begin{equation}
\begin{split}
\frac{\delta p-\left|\delta p\right|M}{1-\left|\delta p\right|M/\left(1-p_c\right)}&=\\
-\frac{1}{\widetilde{z}}\left(\int_0^{1}\textrm{d}\xi\frac{1}{1-\left(\mu/w_{\textsc{L}}\right)\xi}\right.&\left.+\int_0^{1}\textrm{d}\xi\frac{1}{1-\left(\mu/w_{\textsc{T}}\right)\xi}\right).
\end{split}
\end{equation}
Assuming $\textrm{Im}\left(\mu\right)<0$ for $w>0$, we have
\begin{equation}
\begin{split}
\frac{\delta p-\left|\delta p\right|M}{1-\left|\delta p\right|M/\left(1-p_c\right)}&=\\
\frac{1}{\widetilde{z}}\left(\frac{w_{\textsc{L}}}{\mu}\log\left(1-\frac{\mu}{w_{\textsc{L}}}\right)\right.&\left.+\frac{w_{\textsc{T}}}{\mu}\log\left(1-\frac{\mu}{w_{\textsc{T}}}\right)\right)
\end{split}
\end{equation}
We are interested in the low-frequency behavior. There is now quite clearly a logarithmic singularity at low frequencies, as discussed in~\cite{During2013}\footnote{Their parameter $\varepsilon$ is our $-\lim_{\omega\rightarrow0}f$ on the floppy side of the transition with undamped dynamics.}. Keeping only the leading-order low frequency terms, we have
\begin{equation}
\frac{\delta p-\left|\delta p\right|M}{1-\left|\delta p\right|M/\left(1-p_c\right)}\approx -c_1\frac{w}{\mu}\log\left(-c_2\frac{w}{\mu}\right).
\end{equation}
One cannot define a scaling variable for frequency as a ratio of $w$ and $\delta p$ raised to some power as we did in $d=3$, as extra factors of $\delta p$ appear inside the logarithm when written in terms of the scaling variables. We instead implicitly define a scaling variable for $f\equiv w/\mu$ as the right-hand side divided by $\left|\delta p\right|$:
\begin{equation}
\begin{split}
\pm1-M&=F_2,\\
\left|\delta p\right|F_2&\equiv-c_1f\log\left(-c_2f\right).
\end{split}
\end{equation}
The variable $F_2$ is the 2D quantity that corresponds to the right-hand side of Equation~\eqref{eq:scaling3D}. Given that $F_2$ is an invariant scaling combination, we can investigate how this implies $f$ depends upon $\left|\delta p\right|$ by inverting the definition of $F_2$:
\begin{equation}\label{eq:2Dfreq}
f=-\frac{\left(F_2/c_1\right)\left|\delta p\right|}{W_{-1}\left(\left(F_2c_2/c_1\right)\left|\delta p\right|\right)},
\end{equation}
where $W_{-1}(z)$ is a particular branch of the Lambert $W$ function satisfying $W(z)e^{W(z)}=z$. The appropriate branch of the $W$ function has an expansion near $z=0$ of the form $W_{-1}(z)\sim\textrm{log}\left(z\right)-2\pi i-\log\left(\textrm{log}\left(z-2\pi i\right)\right)+\dots$ (see~\cite{LambertW1996Knuth} for more details). This shows directly that the appropriate scaling variable for frequency has important logarithmic shifts close to the critical point in $d=2$. These logarithmic shifts are confirmed in Figure~\ref{TriangularCPANumerics}, which compares our asymptotic scaling forms for the continuum CPA to a numerical solution of the CPA for the bond-diluted triangular lattice. More details of this comparison are located in Appendix~\ref{AppD}.

We will now seek to unify these contrasting results for the scaling variables in 3D and 2D by passing to continuous spatial dimensions $d$ and analyzing the transition as a function of spatial dimension, so that $d=2$ and $d=3$ appear as special cases.

\section{General dimensions: RG flows and Universal Scaling functions}\label{GenDim}

We now extend our analysis in the previous section to arbitrary continuous dimensions $d$. We show that new critical exponents arise for dimensions $d<2$ (Section~\ref{subsec:dBelow2}); we discuss their relevance and possible models to probe these new critical properties in Section~\ref{Applicability}.
We use our predictions to deduce that the critical properties near two dimensions are described by a transcritical bifurcation in the renormalization group flows (as in the 4D Ising model), use normal form theory~\cite{ArchishmanNF} to predict the universal nonlinear terms needed in the RG, and use our exact solutions to deduce the CPA predictions for the values of these universal nonlinear terms (Section~\ref{DeducedRG}). We then investigate the fate of all invariant scaling combinations as we tune through $d=2$ dimensions (Section~\ref{VarsandFuncts}), and write universal scaling functions for the viscoelastic shear modulus in terms of the appropriate scaling variables in each dimension.
All details of the direct calculation from the CPA can be found in Appendix~\ref{AppC}; here we report the crucial parts necessary for the understanding of the invariant scaling combinations and the calculation of scaling functions. Definitions of some variables that regularly appear can be found in Table~\ref{tab:DefinitionsTable}, along with their location in the manuscript.

\begin{table*}[]
    \centering
    \begin{tabular}{|c|c|c|}
    \hline
    Variable & Description & Location \\
    \hline
    $\mu$ & viscoelastic shear modulus & --- \\
    \hline
    $\delta p$ & $p-p_c$: deviation from continuous rigidity transition $p_c$ & --- \\
    \hline
    $w$ & $\rho\omega^2$: (undamped); $i\gamma\omega$: (overdamped) & --- \\
    \hline
    $f$ & $w/\mu$: frequency measured relative to shear stiffness & --- \\
    \hline
    $M$ & $\left(\mu/\mu_0\right)/\left|\delta p\right|$: scaling variable formed between $\mu$ and $\delta p$ & Eqn.~\eqref{eq:MDefinition}, Eqn.~\eqref{eq:MFormula1}, Eqn.~\eqref{eq:MFormula2}\\
    \hline
    $F$ & scaling variable formed between $f$ and $\delta p$ in $d>2$ & Eqn.~\eqref{eq:FDefinition} \\
    \hline
    $F_d$ & scaling variable formed between $f$ and $\delta p$ in $d<2$ & Eqn.~\eqref{eq:FDefinitionLower} \\
    \hline
    $G$ & scaling variable associated with leading irrelevant correction to scaling in $d>2$ & Eqn.~\eqref{eq:GDefinition} \\
    \hline
    $G_d$ & scaling variable associated with leading irrelevant correction to scaling in $d<2$ & Eqn.~\eqref{eq:GDefinitionLower} \\
    \hline
    $\mathcal{D}$ & scaling form of density of states for $d>2$ & Eqn.~\eqref{eq:DDefinition}\\
    \hline
    $\mathcal{D}_d$ & scaling form of density of states for $d<2$ & Eqn.~\eqref{eq:DDefinitionLower}\\
    \hline
    $\mathcal{G}$ & scaling form of the Green's function in $d>2$ & Eqn.~\eqref{eq:GreenDefinition} \\
    \hline
    $\mathcal{G}_d$ & scaling form of the Green's function in $d<2$ & Eqn.~\eqref{eq:GreenDefinitionLower} \\
    \hline
    \end{tabular}
    \caption{\textbf{Descriptions of physical parameters and scaling variables.} The location within the manuscript of the definition of each variable is also included. Note that we use $G$ for the dangerous irrelevant variable
and $\mathcal{G}$ for the Green's function; in the literature $G$ is often used to denote the shear modulus, which we call $\mu$.}
    \label{tab:DefinitionsTable}
\end{table*}

\subsection{New exponents below $d=2$}
\label{subsec:dBelow2}
The CPA self-consistent equation (Equation~\eqref{eq:SC}) depends explicitly on the frequency only through $\textbf{G}$ in the integral term, which splits naturally into a transverse and a longitudinal part, each of which can be evaluated separately. Following the derivation of the scaling in $3$ and $2$ dimensions, we first subtract $p_c$ from each side of the equation, so that, to leading order in the scaling variables, each side of the equation $\sim\left|\delta p\right|^1$. Focusing arbitrarily on the longitudinal part, one can rescale the integration variable to $\xi=(q/q_{\textsc{D}})^2$, to find integrals of the form
\begin{equation}
-\frac{d}{2\widetilde{z}}\int_0^1\textrm{d}\xi\;\frac{\xi^{d/2-1}}{1-(\mu/w_{\textsc{L}})\xi}=-\frac{1}{\widetilde{z}}\,{}_2\textrm{F}_1\left(1,\frac{d}{2};\frac{d}{2}+1;\frac{\mu}{w_{\textsc{L}}}\right),
\end{equation}
where $w_{\textsc{L}}$ depends on the dynamics (overdamped, undamped) but is generally a frequency variable rescaled by longitudinal information; for undamped dynamics $w_{\textsc{L}}=\rho\omega^2/(\lambda_{\textsc{F}}/\mu_{\textsc{F}}+2)q_{\textsc{D}}^2$. The function ${}_2\textrm{F}_1\left(a,b;c;z\right)$ is the ordinary hypergeometric function. Note that $\mu$ is a complex number determined by the solution to the self-consistent equation. In this way, the analytic structure of the asymptotic scaling for $\mu$ is closely linked to the analytic structure of the hypergeometric function, which is convoluted enough to justify further investigation.

We now need to expand the hypergeometric function as $\mu/w_{\textsc{L/T}}$
reaches its limiting value in the scaling limit. It is well-known in the field, and was found by us directly in Section~\ref{subsec:3d}, that this 
ratio diverges in the scaling limit in $d=3$.%
 \footnote{Mean-field rigidity transitions have $\mu \sim |\delta p|$. 
 In the the undamped case $w=\rho \omega^2$ and it
 is well known that $\omega \sim |\delta p|$, so 
 $\mu/w \sim 1/|\delta p|$ diverges.
 In the overdamped case $w = i \gamma \omega$,
 and we shall find $\omega \sim |\delta p|^2$, so again the ratio diverges as 
 $1/|\delta p|$. In fact, our analysis is agnostic to the type of
 damping, so $w_{\textsc{L}/\textsc{T}} \sim |\delta p|^2$ above $d=2$ for 
 all kinds of $q$-independent damping. This tells us the
 scaling of $\omega$ and hence that the ratio diverges.}
We will assume that this is true in all dimensions (without assuming any specific power laws) and self-consistently check it at the end. The hypergeometric function has a branch point at $z=\infty$ that we need to account for to do our expansion properly. There is a relation between the hypergeometric function's values inside the unit disk $\left|z\right|<1$ and those outside that nicely elucidates the complex branch structure at $\infty$. Assuming $d$ is not an even integer~\cite{GRIntegrals}, we can separate our expression into two pieces
\begin{equation}
\begin{split}
\frac{\Gamma\left(\frac{d}{2}\right)^2}{\Gamma\left(\frac{d}{2}+1\right)\Gamma\left(\frac{d}{2}-1\right)}{}_2\textrm{F}_1\left(1,\frac{d}{2};\frac{d}{2}+1;\frac{\mu}{w_{\textsc{L}}}\right)=\\
\left(-\frac{w_{\textsc{L}}}{\mu}\right){}_2\textrm{F}_1\left(1,1-\frac{d}{2};2-\frac{d}{2};\frac{w_{\textsc{L}}}{\mu}\right)+\\
+\Gamma\left(\frac{d}{2}\right)^2\frac{\Gamma\left(1-\frac{d}{2}\right)}{\Gamma\left(\frac{d}{2}-1\right)}\left(-\frac{w_{\textsc{L}}}{\mu}\right)^{d/2}.
\end{split}
\end{equation}
We will eventually recover the behavior in even dimensions $d$ by carefully taking a limit. The hypergeometric function with $z=w_{\textsc{L}}/\mu$ as an argument is $1$ for $z=0$ and can otherwise be expanded in a convergent power series in $z$. In other words, from our self-consistent equation, the longitudinal piece can be expanded in the scaling limit as
\begin{equation}\label{eq:Freqscaling}
\begin{split}
-\frac{1}{\widetilde{z}}\,{}_2\textrm{F}_1\left(1,\frac{d}{2};\frac{d}{2}+1;\frac{\mu}{w_{\textsc{L}}}\right)&=\\
C_1\left(\frac{w_{\textsc{L}}}{\mu}\right)+C_2\left(-\frac{w_{\textsc{L}}}{\mu}\right)^{d/2}+&\mathcal{O}\left(\left(\frac{w_{\textsc{L}}}{\mu}\right)^2\right).
\end{split}
\end{equation} 
In the previously analyzed $d=3$, for instance, this identifies the correction fixing the low-frequency imaginary part of the modulus as a non-analytic $\left(-w_{\textsc{L}}/\mu\right)^{3/2}$ appearing in the self-consistent equation -- related to our dangerously irrelevant variable.

One now tries to set the scaling of $w_{\textsc{L}}/\mu$ by examining Equation~\eqref{eq:Freqscaling}.
For $d>2$, the term $\left(-w_{\textsc{L}}/\mu\right)^{d/2}$ is subdominant at low frequencies to $w_{\textsc{L}}/\mu$, and so $w_{\textsc{L}}/\mu\sim\left|\delta p\right|$ so that either side of the self-consistent Equation~\eqref{eq:SC} is balanced asymptotically as $\delta p\rightarrow 0$. On the other hand, for $d<2$, $\left(-w_{\textsc{L}}/\mu\right)^{d/2}$ sets the dominant contribution at low frequencies. In this scaling limit, $w_{\textsc{L}}/\mu\sim\left|\delta p\right|^{2/d}$ so that $\left(-w_{\textsc{L}}/\mu\right)^{d/2}\sim\left|\delta p\right|$ and either side of the self-consistent equation is balanced asymptotically as $\delta p\rightarrow 0$. We note that each of these scalings is consistent with $\mu/w_{\textsc{L}}\rightarrow\infty$ in the scaling limit, justifying the expansion of the hypergeometric function around its branch point at $\infty$. This difference in scalings is reminiscent of a theory near its upper critical dimension: above $d=2$, the critical exponents are dimension-independent and equal to their mean-field-like value. Below $d=2$, the exponents are modified because the mean-field fixed point becomes unstable under the RG flow.

\subsection{Deduced RG flow equations}
\label{DeducedRG}
In our case, we have access not to a principled set of renormalization group transformations for rigidity percolation or jamming, but only to previously studied scaling exponents and our explicit solutions. Here we posit RG flow equations that accurately reproduce the power-law invariant scaling combinations in our explicit solutions (see Appendix~\ref{AppA}). These flow equations are nonlinear functions of the system parameters that express the amount they change as the system is coarse-grained by a factor $1+\textrm{d}\ell$ and rescaled. We find from our explicit solution that we should measure $w$ in units of $\mu$ (measuring frequency in units of stiffness). We also find there is an additional control variable $g$ whose flow depends upon the dimension $d$: irrelevant for $d>2$, becoming marginal in $d=2$, and relevant at the original RG fixed point for $d<2$.

Our group has developed~\cite{ArchishmanNF} an understanding of how theories of critical phenomena behave in the vicinity of their upper and lower critical dimensions. Normal form theory, adapted from dynamical systems, gives a unifying description of renormalization group flows and special invariant scaling combinations. In particular, it gives us the language necessary to interpret flows near a bifurcation. In the Ising model near its upper critical dimension $d=4$, for instance, a variable that can be identified with the quartic coupling $g$ which was once irrelevant in $d>4$ becomes marginal in $d=4$ and relevant below (at the Gaussian fixed point), redirecting the flows to the new, stable Wilson-Fisher fixed point. In the normal form language, the RG flow of this parameter undergoes a transcritical bifurcation around $d=4$. Through an analytic change of coordinates, the flow equations can be cast into their \textit{normal form} in the vicinity of $d=4$. In typical cases (such as in the three-dimensional Ising model), the normal form can completely linearize the flow at the fixed point, giving invariant scaling combinations that are ratios of powers of physical quantities, and hence power-law behavior that can be characterized by critical exponents. But precisely \textit{at} the bifurcation in the upper critical dimension, one must keep specific nonlinear terms which cannot be removed by an analytic change of variables, and these nonlinear terms capture completely the well-known logarithmic corrections.

Based on our explicit solution, we thus posit the following RG flow equations%
\footnote{Normal form theory~\cite{ArchishmanNF} demands a cubic term $D g^3$ in the equation for $\textrm{d}g/\textrm{d}\ell$. We have checked that the constant $D=0$ for our explicit solution.}, which accurately reproduce the power-law invariant scaling combinations in all dimensions $d\neq 2$:
\begin{equation}\label{eq:RGflows}
\begin{split}
\frac{\textrm{d}q}{\textrm{d}\ell}&=q,\\
\frac{\textrm{d}\delta p}{\textrm{d}\ell}&=2\delta p-g\,\delta p,\\
\frac{\textrm{d}\mu}{\textrm{d}\ell}&=2\mu-g\,\mu,\\
\frac{\textrm{d}f}{\textrm{d}\ell}&=2f,\\
\frac{\textrm{d}g}{\textrm{d}\ell}&=(2-d)g-g^2,
\end{split}
\end{equation}
where the nonlinear terms can be removed in dimensions $d>2$. Details of the determination of these flow equations can be found in Appendices~\ref{AppA} and~\ref{AppB}.
In the flow equations for parameters other than $g$, the terms involving products of flow parameters are higher-order and so the invariant scaling combinations in dimensions other than $d=2$ can be accurately determined by setting $g$ equal to its value at the stable RG fixed point, as we will do in the following section. 

\subsection{Scaling variables and scaling functions}
\label{VarsandFuncts}

\begin{figure}[ht]
\begin{center}
\includegraphics[width=0.9\linewidth]
{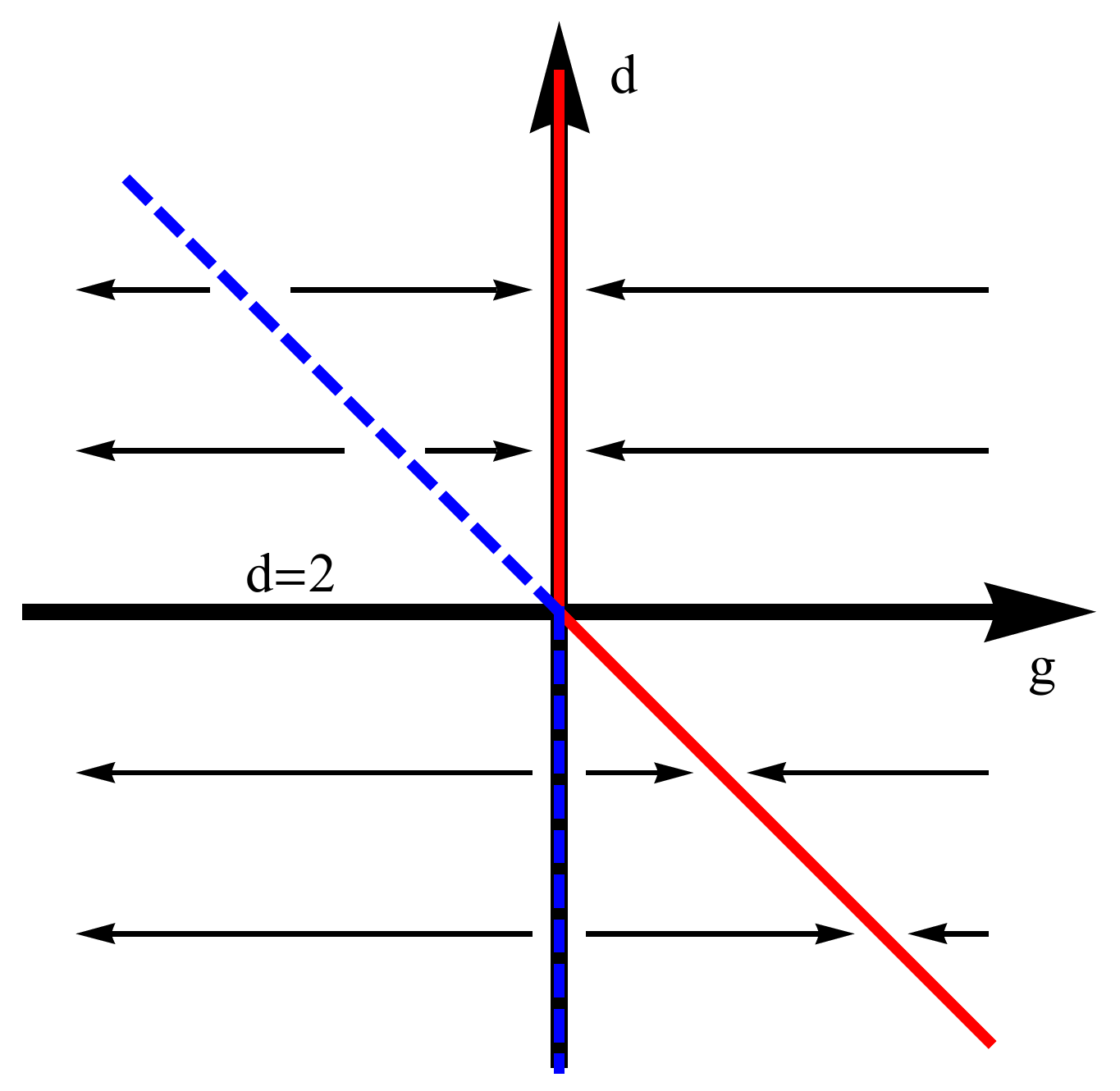}
\end{center}
\caption{\textbf{Flows of the parameter $g$ under the RG}. The solid, red lines indicate attractive fixed points. The dashed, blue lines indicate repulsive fixed points. For $d>2$, $g$ flows to $0$. For $d<2$, $g$ flows to $g_c=2-d$, a different fixed point. This is a transcritical bifurcation; two fixed points cross and exchange their stability in $d=2$. The dimension-dependent critical exponents for $d<2$ are controlled by the new stable fixed point.}
\label{gRGFlows}
\end{figure}

We now use the flow equations deduced from the scaling behavior together with the asymptotic expansions of the CPA close to the critical point to write scaling variables and scaling functions for the viscoelastic moduli in arbitrary dimensions. Turning our focus to the flow equation for $g$, we find $g_c=0$ for $d>2$ and $g_c=2-d$ for $d<2$: by construction, the flow equation for $g$ undergoes a transcritical bifurcation in $d=2$ (Figure~\ref{gRGFlows}).

To demonstrate that the scaling for $f$ deduced from the flow equations is consistent with the direct CPA calculation both above and below $d=2$, we combine the flow equations for $\delta p$ and $f$ to write
\begin{equation}
\frac{\textrm{d}\log\left(f\right)}{\textrm{d}\log\left(\delta p\right)}\approx\frac{2}{2-g_c}\implies f\sim\delta p^{2/(2-g_c)}.
\end{equation}
This is correct away from $d=2$, as $g_c=0$ makes the exponent $2/(2-g_c)=1$ and $g_c=2-d$ makes the exponent $2/(2-g_c)=2/d$. (We will explore $d=2$ below, where the full flow equations including the nonlinear  terms are needed to determine the complicated invariant scaling combinations.) For $d>2$, we must retain the flow equation for $g$ despite its irrelevance; it is a dangerously irrelevant variable since the scaling function for $M$ is not analytic at $G=0$.

Is this interesting, given that we typically don't do simulations of colloidal gels or other disordered rigid systems \textit{below} two dimensions? 
While we do not expect experiments to test our predictions in non-integer dimensions, in Section~\ref{Applicability} we discuss applications to theoretical simulations in $d<2$. In two dimensions, however, these predictions
are in principle experimentally observable.
This is a case where the upper critical dimension of a theory is physically relevant (at least as determined within the CPA). In the Ising model at its upper critical $d=4$, there is a logarithmic singularity in the magnetization, which also has important (\textit{very} slowly vanishing at the critical point) log-log corrections. Here, the situation is identical. Suppose one picks out a physically relevant piece of dynamical information at this rigidity transition, like the frequency $\omega^*$ where we cross over into being dominated by dissipation $\textrm{Re}\left(\mu\left(\omega^*\right)\right)=-\textrm{Im}\left(\mu\left(\omega^*\right)\right)$. For all $d$ away from $2$, we can write how this physical frequency scales with our excess contact number: in the undamped case, $\omega^*\sim\delta p^{(4-g_c)/(4-2g_c)}$. But in 2 dimensions, there are detectable shifts in this frequency, and it turns out that for the undamped case $\omega^*\sim\left|\delta p\right|\left|\log\left|\delta p\right|\right|^{-1/2}$ with additional important log-log corrections.

This can be understood by writing the proper invariant scaling combinations in the upper critical dimension using the flow equations. In Appendix~\ref{AppB}, we show that the scaling of the frequency variable in 2 dimensions implied by the renormalization group flow equations is (making only the $\delta p$ dependence explicit)
\begin{equation}
f\sim\frac{x(g)\delta p}{W\left(x(g)\,\delta p\right)},
\end{equation}
where $W$ is again the Lambert $W$ function and $x(g)$ is an explicit function of the (marginally) irrelevant scaling variable. This accurately reproduces the asymptotic behavior of the frequency scaling that was found by directly evaluating the CPA in $d=2$ (Equation~\eqref{eq:2Dfreq}) and connects this result to scaling found in standard critical phenomena.

As investigated in the particular cases of $d=3$ and $d=2$ in previous sections, the CPA self-consistent equation also gives us predictions for the universal scaling \textit{functions} determining the moduli; when written in the appropriate scaling variables for each dimension one can predict the shape of scaling collapse plots of measurements of viscoelastic moduli close to a rigidity transition. With this in mind, we seek to write the self-consistent equations in a form that is consistent with the posited renormalization group flow equations.

For dimensions $d>2$ not even, retaining the lowest-order terms leads to a scaling function of the form
\begin{equation}\label{eq:MFormula1}
\begin{split}
\pm1-M&=F+G\left(-F\right)^{d/2},\\
F&=\frac{f/f_0}{\left|\delta p\right|},\;
M=\frac{\mu/\mu_0}{\left|\delta p\right|},\;G=g/g_0\left|\delta p\right|^{d/2-1},
\end{split}
\end{equation}
where the scaling variable $F$ is the invariant scaling combination formed between $f$ and $\delta p$ and $f_0$, $\mu_0$, and $g/g_0$ are complicated but explicit combinations of microscopic parameters (Appendix~\ref{AppC}). The sign $\pm1=\delta p/\left|\delta p\right|$ is for the rigid and floppy side of the transition, respectively. The scaling variable $F$ can be adapted to deal with different microscopic dynamics. Here we note that the scaling variable $G$ varies with the correct power of $\delta p$ to be identified as the invariant scaling combination formed between $g$ and $\delta p$. Thus $g$ is correctly identified as being irrelevant in dimension $d>2$. As mentioned before in Section~\ref{23Dim}, in previous work \cite{LiarteTSCCS22,LiarteTSCCSnn2}, $G$ is set to $0$, making the eventual solution for $\mu(\omega)$ the solution to a quadratic equation, with a cuspy form not seen in the full numerical solution to the CPA. The variable $g$ is dangerously irrelevant for microscopically undamped dynamics in all $d>2$ and needs to be retained to understand the behavior of the low-frequency viscous part of the modulus. We also note that the effect of $G$ is nonperturbative in the inverse spatial dimension: near $d=\infty$ perturbing in $\epsilon\equiv1/d$, $G\sim\left|\delta p\right|^{1/(2\epsilon)}$. This effect is not specific to the CPA, since above the upper critical dimension the onset of the plateau in the density of states is $\omega^*\sim\left|\delta p\right|$ while a description of the phononic contribution to the low-frequency density of states gives Debye behavior $\sim\omega^{d-1}$ which is heavily suppressed in high dimensions.

Below dimension $2$, we are forced to change our scaling variables to the ones relevant at the new attractive RG fixed point. We find

\begin{equation}\label{eq:MFormula2}
\begin{split}
\pm1-M&=-\left(-F_d\right)^{d/2}-G_{d}F_d,\\
F_{d}&=\frac{f/f_{0d}}{\left|\delta p\right|^{2/d}},\;G_{d}=g_d/g_{0d}\left|\delta p\right|^{2/d-1}.
\end{split}
\end{equation}
This identifies $G_d$ as the invariant scaling combination associated with $g$ in the vicinity of the new attractive fixed point that has emerged below $d=2$, with the correct exponent on $\delta p$ (see Appendix~\ref{AppA} and Appendix~\ref{AppC} for details). We summarize the usual critical exponents in Table~\ref{tab:CritExp}.

\begin{table}[]
    \centering
    \begin{tabular}{|c|c|c|c|c|c|c|}
    \hline
     & $f_{\textrm{CPA}}$ & $\nu$ & $z$ & $\theta\equiv\omega\nu$ & $\gamma$  \\
     \hline
     \hspace{0.1cm} $d>2$ \hspace{0.1cm} & \hspace{0.1cm} $1$ \hspace{0.1cm} & \hspace{0.1cm} $1/2$ \hspace{0.1cm} & $2$ & $d/2-1$ & $2$ \\
     \hline
     $d<2$ & $1$ & $1/d$ & \hspace{0.1cm} $1+d/2$ \hspace{0.1cm} & \hspace{0.1cm} $2/d-1$ \hspace{0.1cm} & \hspace{0.1cm} $1+2/d$ \hspace{0.1cm} \\
     \hline
    \end{tabular}
    \caption{\textbf{Critical exponents as predicted by the CPA} in the undamped case, away from the upper critical dimension ${d_\mathrm{upper}=2}$. The invariant scaling combinations in ${d_\mathrm{upper}}$ capture the logarithmic corrections typical of an upper critical dimension.}
    \label{tab:CritExp}
\end{table}

One can recover the scaling function in $d=2$ by carefully taking a limit $d\rightarrow 2$. Reinstalling the definition of $G$ above $d=2$, and pulling out a factor of $\Gamma\left(d/2-1\right)$ that was previously absorbed into a definition of $f_0$, one finds
\begin{equation}
\begin{split}
\pm1-M&=\\
-F'&\left(\left(-(g/g_0)'^{\frac{2}{d-2}}F'\left|\delta p\right|\right)^{d/2-1}-1\right)\Gamma\left(\frac{d}{2}-1\right).
\end{split}
\end{equation}
As $d\rightarrow2^+$, this reproduces a limit similar to those seen in the replica trick $(x^n-1)/n\rightarrow\log(x)$ as $n\rightarrow 0$ (elaborated upon in Appendix~\ref{AppC}). The resulting expression
\begin{equation}
\pm1-M=-F'\log\left(-(g_2/g_{02})F'\left|\delta p\right|\right)
\end{equation}
gives the right asymptotic behavior predicted by the RG flows \textit{at} the transcritical bifurcation, but is not written in the appropriate scaling variables for $d=2$. One can see this because if one fixes these scaling variables but takes the scaling limit $\delta p\rightarrow0$, one side of the self-consistent equation diverges. If one inserts the proper frequency variable, the equation is asymptotically
\begin{equation}\label{eq:2DUniversalSFM}
\begin{split}
\pm1-M&=F_2,\\
F_2&=\frac{f/f_{02}}{\left|\delta p\right|}W\left(x(g)\,\delta p\right).
\end{split}
\end{equation}
Explicit formulas for all quantities in terms of the microscopic parameters of the isotropic elastic sheet are also derived in Appendix~\ref{AppC}. In Figure~\ref{TriangularCPANumerics}, we numerically solve the CPA self-consistent equations for the triangular lattice in the undamped case, which is not microscopically isotropic and which has a hexagonal Brillouin zone. Nonetheless, because there is an emergent long-wavelength isotropy, the scaling behavior of its modulus near the critical point is well-described by our forms of the scaling function for the isotropic elastic sheet. Performing the na\"ive rescalings 
$M_{\textsc{T}}=\mu/\left|\delta p\right|$ and $\Omega_{\textsc{T}}=\omega/\left|\delta p\right|$, one sees a slow but systematic shift of the crossing point $\Omega_{\textsc{T}}^*$ where $\textrm{Re}\left(M_{\textsc{T}}(\Omega_{\textsc{T}}^*)\right)=-\textrm{Im}\left(M_{\textsc{T}}(\Omega_{\textsc{T}}^*)\right)$. If $\Omega_{\textsc{T}}$ were the correct scaling variable to use, as it is in all $d>2$, this crossing point would have small corrections away from $p=p_c$ but would otherwise be constant in $\Omega_{\textsc{T}}$. Our form of the universal scaling function in $d=2$ perfectly accounts for these logarithmic shifts.

\section{Applicability of the transition to physical systems}\label{Applicability}

To what systems do we expect the behavior to be quantitatively described by our universal scaling predictions? What features of our predictions do we expect to apply more broadly to rigidity transitions? Here we discuss clearly where
our universal predictions do not apply, and speculate about where they may provide qualitative
or quantitative guidance.

As noted in the introduction and in Appendix~\ref{AppD}, our continuum
CPA does not correctly describe the diluted triangular spring lattice,
which exhibits non mean-field critical exponents in its static properties in two dimensions rather
than the predicted log-corrections to mean-field
theory~\cite{BroederszMac2011, WillWIP}. Generalized
to describe the abrupt jump in the bulk modulus, the continuum CPA applies qualitatively to 
simulations of spring network geometries generated from jammed
packings~\cite{GoodrichLiuLogsPRE2014}. The generalized continuum CPA reproduces the numerically observed jamming mean-field exponents in three dimensions, and both this calculation and the simulation find
mean-field exponents with log corrections to scaling in two dimensions
(however, see below). Indeed, based partly on a discussion years ago with
Carl Goodrich, we conjecture that randomly diluting a spring
network generated from a jammed packing starting in the rigid phase will
undergo a transition where both bulk and shear moduli grow continuously,
described quantitatively by the version of our continuum CPA analyzed here~\cite{GoodrichPrivateCommunication}.

Why is the triangular lattice (and, by implication, many other spring lattices)
not behaving according to our theory? 
While the triangular lattice is statistically isotropic in its elastic 
moduli, many critical points are more sensitive than elastic theory to
breaking of rotational invariance. The XY model is unstable to breaking
of triangular and square symmetries~\cite{XY4foldTheory,XY4foldExperiment},
and diffusion-limited aggregation is famous for breaking rotational invariance
in a way that revealed itself only in (then) large
simulations~\cite{DiffusionLimitedAggregation}. Statistically
isotropic lattices might show mean-field behavior. Another likely culprit
are the straight lines between bonds in the triangular lattice, which are 
shared with the Mikado networks~\cite{FredReview2014} formed by random long fibers
cross-linked at their intersections. A node connecting two parallel bonds
cannot move freely under tensile stress, while at any non-zero angle it needs
a third constraint to fix it in place: such \textit{second-order constraints}
are expected to change critical properties~\cite{jacobs1995generic,ManningSantangelo}. A generic lattice with the same connectivity structure as a regular triangular lattice, as suggested by 
Jacobs~\cite{jacobs1995generic}, could be enough to show mean-field behavior. And diluted, statistically isotropic spring networks with random node positions show mean-field exponents~\cite{DeGiuli2014}.

As noted above, this continuum analysis has a natural extension to describe a jamming transition, inspired again by~\cite{LiarteXOL19, LiarteTSCCS22, LiarteTSCCSnn2}. There the upper critical dimension is believed to be two, and our model (and the lattice CPA models) does agree with known critical exponents. However, our calculation for jamming shows logarithmic corrections to scaling that only arise in frequency-dependent quantities, as we have seen in this manuscript, while convincing numerical work shows logarithmic corrections to the finite-size scaling of zero-frequency elastic moduli in jamming simulations~\cite{GoodrichLiuLogsPRE2014}. 
We suspect that the lack of logarithmic corrections to the finite-size effects in our 2D jamming variant is related to the CPA assumption that the moduli are independent of wavevector.
One notes that the Green's function and other wavevector-dependent properties  have a wavevector scale in 2D that does have logarithmic corrections to scaling $q^*\sim\left|\delta p\right|^{1/2}\left|\log\left|\delta p\right|\right|^{-1/2}$ (similar to the frequency scaling).
Our current theory replaces the system near its rigidity transition with a uniform but
frequency dependent modulus, ignoring the important effects of spatial fluctuations.
A more sophisticated self-consistent approach allowing for more general forms of response functions, such as the preprint of Vogel et al.~\cite{Fuchs2024arxiv} mentioned in the Introduction, could be analyzed to extract 
a universal scaling theory that not only correctly describes the finite-size effects in two-dimensional jamming, but also describes the 2D and 3D crossovers to the elastic-dipole induced correlations of glasses~\cite{MaierZF18,ZippeliusSethnaTextbook2023,BulbulGranularStatMech2009} and Rayleigh scattering. However, the detailed simulations of~\cite{During2013} do appear to reproduce the logarithmic shifts in the density of states and $\ell_c^{-1}\sim q^*$ explained in this work.

Will our predictions of new critical behavior for $d<2$ in Section~\ref{GenDim} have any chance of being tested? Jamming has been studied for hard disks in a one-dimensional channel~\cite{ZhangGM20}, exhibiting different critical exponents for the distribution of small gaps than in higher dimensions. We suspect our predicted critical exponents for linear response properties may be only qualitatively predictive in $d=1$. While our theory predicts exact new values for critical exponents for all $d<d_\mathrm{upper} = 2$, we trust our predictions quantitatively only near the upper critical dimension. Indeed, other approximate methods such as mode-coupling theories~\cite{HohenbergPC,HohHalp1977} give correct critical exponents in $d_\mathrm{upper}-\epsilon$ only to order $\epsilon^1$, where higher-order corrections demand further diagrammatic calculations. 

The calculation presented here could, however, quantitatively describe a rigidity transition in a one-dimensional model with long-range bonds. For instance, the one-dimensional Ising model~\cite{AngeliniPR-T14} with bond strengths that decay with a power law $J(r) \sim r^{-(1+\sigma)}$ has an ordering transition for $\sigma<\sigma_{\textsc{L}}=1$ and has nontrivial critical exponents for $1/2<\sigma<1$. The situation is similar in ordinary percolation with long-range bonds~\cite{GoriPRE2017}, where having a bond length distribution $P(r)\sim r^{-(1+\sigma)}$ leads to a percolation transition at a threshold $0<p_c<1$ for $0<\sigma<1$, and nontrivial critical exponents for $1/3<\sigma<1$ with additional logarithmic corrections to scaling at $\sigma=1/3$. This could be extended to an elastic rigidity transition by replacing the connecting bonds by elastic springs and measuring the elastic modulus. Additionally, the long-range random-bond Ising model in one dimension has a spin glass phase~\cite{KotliarAS83}. These are all cases where tuning the exponent associated with the range of the interaction can be used to continuously tune the effective dimension of the critical properties of the transition, allowing us to access the continuous predictions of critical exponents in the vicinity of the upper critical dimension.

Finally, additional extensions to this scaling theory can in principle be added by hand. Exactly \textit{at} the rigidity transition, there is no linear response regime, which has attracted much interest in the jamming community. Tiny deformations induce both microscopic topology changes and avalanches of all sizes. Scaling theories have been developed to describe, for instance, power-law relationships between friction coefficients and shear rate in granular matter close to its flowing instability~\cite{ZippeliusSethnaTextbook2023,PerrinWyartMetzger2021}. Analytical predictions for the universal scaling functions may not be as simple to determine, but once a suitable scaling theory is established, adding new phenomena (such as rheological responses beyond the linear regime) should be possible.

\section{Summary \& Conclusions}\label{Conclusion}
In summary, we take seriously a dynamical version of the CPA, a frequently used effective medium theory. We examine its predictions close to the critical point, casting solutions into scaling forms to identify universal pieces. We first examine its predictions for the universal scaling functions for effective viscoelastic moduli close to the critical point in $d=3$ and identify a dangerously irrelevant variable that controls low-frequency dissipation in the case of microscopically undamped dynamics. We then investigate $d=2$, and we find that the appropriate invariant scaling combinations are not ratios of powers of parameters, as we would expect close to a hyperbolic RG fixed point. To our surprise, although the exponents with which the \textit{static} moduli vanish with $\delta p$ are unchanged with dimension, the critical exponents associated with relevant length and time scales in the system change quantitatively as we pass below $d=2$. This identifies $d=2$ as the upper critical dimension of the theory. From the exact solution, we deduce from normal form theory~\cite{ArchishmanNF} a set of renormalization group flow equations which have a transcritical bifurcation in two dimensions. These are constructed to match the forms of the scaling variables above, below, and in the upper critical dimension $d=2$. These forms are self-consistently checked against numerical solutions of the lattice CPA for a bond-diluted triangular lattice, verifying these important corrections in this physically relevant dimension.

\acknowledgments
The authors thank Gilles Tarjus for recommending deeper analysis of the connection with the $4-\epsilon$ expansion of the Ising model. SJT thanks Bulbul Chakraborty for interesting discussions about logarithmic corrections in the upper critical dimension. SJT, IC, and JPS are supported by NSF DMR-2327094. DBL is supported by FAPESP through grants \# 2021/14285-3 and \# 2022/09615-7. 

\appendix

\section{Renormalization Group flows from scaling combinations}\label{AppA}
Here we detail the procedure of writing down our deduced renormalization-group flow equations, (Section~\ref{DeducedRG}). First, we take for granted that the wavevector, as an inverse length, coarse grains as
\begin{equation}
\frac{\textrm{d}q}{\textrm{d}\ell}=q.
\end{equation}
This can be taken as a definition of the coarse-graining procedure. Let us first focus on flows above the upper critical dimension and ignore the irrelevant variable. The frequency variables scale as $f\equiv w/\mu\sim\delta p$, and the modulus scales as $\mu\sim\delta p$. From the form of the Green's function, $\left(\mu q^2-\mu f\right)^{-1}\sim \delta p^{-\gamma}$, which means that $q\sim\delta p^{1/2}$ gives a nontrivial scaling limit. These mean-field power laws determine the coefficients on the linear parts of the flow equations, i.e. the terms linearized about the hyperbolic fixed point above $d=2$. We have for $d>2$ the normal form:
\begin{equation}
\begin{split}
\frac{\textrm{d}q}{\textrm{d}\ell}&=q,\\
\frac{\textrm{d}\delta p}{\textrm{d}\ell}&=2\delta p,\\
\frac{\textrm{d}\mu}{\textrm{d}\ell}&=2\mu,\\
\frac{\textrm{d}f}{\textrm{d}\ell}&=2f.
\end{split}
\end{equation}

Now we include the effects of an upper critical dimension of $2$. Assuming that a variable that was originally irrelevant undergoes a transcritical bifurcation and becomes relevant below $d=2$, we can write the flow equations with help from normal form theory~\cite{ArchishmanNF}. Normal form theory tells us the minimal number of terms we need to keep assuming we have made an analytic change of coordinates, i.e., we have preserved information about the singularity close to the critical point. We write a flow equation for this speculative coupling $g$ irrelevant above $2$ dimensions that undergoes a bifurcation and flows to a new stable point for $d<2$. We must keep terms linear in $g$ that appear in the other flow equations; these serve to modify the critical exponents below the upper critical dimension. We have

\begin{equation}
\begin{split}
\frac{\textrm{d}q}{\textrm{d}\ell}&=q,\\
\frac{\textrm{d}\delta p}{\textrm{d}\ell}&=2\delta p-\lambda_p g\,\delta p,\\
\frac{\textrm{d}\mu}{\textrm{d}\ell}&=2\mu-\lambda_\mu g\,\mu,\\
\frac{\textrm{d}f}{\textrm{d}\ell}&=2f-\lambda_f g\,f,\\
\frac{\textrm{d}g}{\textrm{d}\ell}&=\frac{1}{A}\left(2-d\right)g-g^2
\end{split}
\end{equation}
where $A>0$ and we rescale $g$ to set the coefficient of the quadratic term in its flow equation to $-1$. As this theory is simple and exactly solvable, we have many \textit{resonances} where there are integer relationships between coefficients in the RG flow equations. We will ignore these for now, but normal form theory gives a prescription to keep additional terms in the flow equations. These are then tuned to capture what would otherwise be interpreted as large corrections to scaling. We check directly that the coefficient $D$ on the cubic term of the flow for $g$, $+Dg^3$, is $0$ through a calculation similar to the one performed in Appendix~\ref{AppB}. The critical value of $g$ below $d=2$ is $g_c=\left(2-d\right)/A$. We know that, below $d=2$, $f\sim\delta p^{2/d}$ and so $q\sim \delta p^{1/d}$. After coarse-graining for a while, we can set $g=g_c$ in the flow equations to get accurate exponents for the invariant scaling combinations. We write
\begin{equation}
\begin{split}
\frac{\textrm{d}\log q}{\textrm{d}\log\delta p}&=\frac{1}{2-\lambda_pg_c}=\frac{1}{2-\lambda_p\left(2-d\right)/A}=\frac{1}{d},\\
\frac{\textrm{d}\log\mu}{\textrm{d}\log\delta p}&=\frac{2-\lambda_\mu g_c}{2-\lambda_pg_c}=\frac{2-\lambda_\mu\left(2-d\right)/A}{2-\lambda_p\left(2-d\right)/A}=1,\\
\frac{\textrm{d}\log f}{\textrm{d}\log\delta p}&=\frac{2-\lambda_f g_c}{2-\lambda_pg_c}=\frac{2-\lambda_f\left(2-d\right)/A}{2-\lambda_p\left(2-d\right)/A}=\frac{2}{d}.\\
\end{split}
\end{equation}
These are solved by
\begin{equation}
\lambda_\mu=\lambda_p=A,\;\;\lambda_f=0.
\end{equation}
We now look at how $g$ scales with $\delta p$: above $d=2$, we have 
\begin{equation}
\frac{\textrm{d}\log g}{\textrm{d}\log\delta p}=\frac{\left(2-d\right)}{2A}=-\frac{1}{A}\left(\frac{d}{2}-1\right).
\end{equation}
Below $d=2$, we have (expanding about the stable RG fixed point $\delta g=g-g_c$)
\begin{equation}
\frac{\textrm{d}\log \delta g}{\textrm{d}\log\delta p}=-\frac{\left(2-d\right)}{dA}=-\frac{1}{A}\left(\frac{2}{d}-1\right).
\end{equation}
The positive constant $A$ sets the scale of $g_c$ for $d<2$. We will determine $A=1$ by looking at the scaling implied by the RG flow equations in $2$ dimensions and choosing $A$ to match the asymptotic scaling found for the frequency directly in $d=2$ from the CPA.
\section{Scaling for frequency in 2 dimensions}\label{AppB}
Here we derive the scaling of $f$ with $\delta p$ in the upper critical dimension $d=2$; other scalings (such as the one for $q$) follow similarly. This closely follows the Supplemental Material of~\cite{ArchishmanNF}. We will use \begin{equation}
\begin{split}
\frac{\textrm{d}g}{\textrm{d}\ell}&=-g^2, \\
\frac{\textrm{d}\delta p}{\textrm{d}\ell}&=2\delta p -A\,g\,\delta p,\\
\frac{\textrm{d}f}{\textrm{d}\ell}&=2f.
\end{split}
\end{equation}
Divide the flow equation for $\delta p$ by the flow equation for $g$ to find
\begin{equation}
\frac{\textrm{d}\delta p}{\textrm{d}g}=\frac{2\delta p-Ag\delta p}{-g^2}.
\end{equation}
This is integrated to give
\begin{equation}
\log\left(\frac{\delta p}{\delta p_0}\right)=2\left(\frac{1}{g}-\frac{1}{g_0}\right)+A\log\left(\frac{g}{g_0}\right).
\end{equation}
It is useful to define a variable
\begin{equation}
s\equiv\frac{1}{g}.
\end{equation}
We coarse-grain to $\delta p=1$. Then
\begin{equation}
-\log\left(\delta p_0\right)=2\left(s-s_0\right)-A\log\left(\frac{s}{s_0}\right).
\end{equation}
Rearrange this into a particular form:
\begin{equation}
\begin{split}
-\frac{2}{A}s+\log\left(-\frac{2}{A}s\right)=\log\left(y\left(s_0\right)\delta p_0^{1/A}\right),\\
y(s_0)\equiv-\frac{2}{A}s_0\,\textrm{exp}\left(-\frac{2}{A}s_0\right).
\end{split}
\end{equation}
Note that this form is one of the definitions of the Lambert $W$ function,
\begin{equation}
W(z)+\log\left(W(z)\right)=\log(z).
\end{equation}
Hence,
\begin{equation}
-\frac{2}{A}s=W\left(y(s_0)\delta p_0^{1/A}\right)
\end{equation}
or
\begin{equation}
s=-\frac{A}{2}W\left(y(s_0)\delta p_0^{1/A}\right).
\end{equation}
Now we use the flow equation for $g$ directly, coarse-graining from $\ell_0=0$ to $\ell$:
\begin{equation}
\ell = \frac{1}{g}-\frac{1}{g_0}=s-s_0.
\end{equation}
We are now prepared to determine how $f_0$ varies with $\delta p_0$ in the upper critical dimension. Directly integrating the flow equation for $f$, we find
\begin{equation}
\frac{f}{f_0}=e^{2\ell}=e^{2\left(s-s_0\right)}=\frac{e^{2s}}{e^{2s_0}}.
\end{equation}
All $\delta p_0$ dependence is through $s$ (not $s_0$). So to find our invariant scaling combination (or the functional dependence of $f_0$ on $\delta p_0$), we write
\begin{equation}
f_0\sim e^{-2s}.
\end{equation}
Inserting our functional form for $s$, using the relation
\begin{equation}
e^{a\,W(z)}=\frac{z^a}{W(z)^a},
\end{equation}
and calling the argument of the $W$ function $z$, we have
\begin{equation}
f_0\sim e^{A W(z)}\sim\frac{z^A}{W(z)^{A}}.
\end{equation}
(We assume $A$ is an integer, which we will see in a moment, to ignore further branch subtleties). Reinstalling the definition of $z$, and calling $x(g_0)=y(s_0)$, we have
\begin{equation}
f_0\sim\frac{\left(x(g_0)\delta p_0^{1/A}\right)^{A}}{W\left(x(g_0)\delta p_0^{1/A}\right)^A}=\frac{x(g_0)^A\,\delta p_0}{W\left(x(g_0)\delta p_0^{1/A}\right)^A},
\end{equation}
Comparing this with the scaling of $f$ determined from the asymptotics of the theory in $d=2$ allows us to identify $A=1$. This gives the result for the flow equations and the invariant scaling combination in the main text. Note that $f\sim\delta p$ (as it does in $d>2$) with additional log and log-log corrections coming from the $W$ function.
\section{Scaling variables in terms of isotropic elastic sheet parameters}\label{AppC}
Here we expound upon the calculation mentioned in the main text that eventually leads to the universal scaling functions for the vanishing modulus. We take the perspective of a long-wavelength limit of a bond percolation CPA that is explicitly isotropic at long-wavelengths; this recovers the CPA for amorphous spring networks developed in~\cite{During2013}. First, we write the dynamical matrix and Green's function for an isotropic elastic sheet, with Lam\'e parameters $\lambda$ and $\mu$. This is typically done in the reverse order using the constitutive relations (i.e. coarse-graining a lattice system with dynamical matrix $\mathbf{D}$), recovering the long-wavelength stiffness tensor $K_{ijk\ell}$. For a lattice with real-space unit cell volume $V$ and $n$ sites in the unit cell, we have
\begin{equation}
K_{ijk\ell}=\frac{n}{V}\left[\frac{1}{2}\frac{\partial^2\textbf{D}^{\textsc{SS}}}{\partial q_j\partial q_{\ell}}-\frac{\partial \textbf{D}^{\textsc{SF}}}{\partial q_j}\left(\textbf{D}^{\textsc{FF}}\right)^{-1}\frac{\partial \textbf{D}^{\textsc{FS}}}{\partial q_{\ell}}\right]_{ik}\Bigg|_{\mathbf{q}=\mathbf{0}},
\end{equation}
where the dynamical matrix is written in the center-of-mass basis and the labels $\textrm{S}$ and $\textrm{F}$ represent the ``slow'' and ``fast'' normal modes~\cite{SoftMatterText} (essentially separating the acoustic modes from the optical). We are interested mainly in a scaling that picks out low-frequency behavior, so we focus on the case $n=1$ with no fast modes (investigating how the more general form interacts with the CPA is a fruitful avenue for future work, and is necessary to fully understand the jamming behavior of e.g. the system studied in \cite{LiarteXOL19}). In this case, the relation simply reads
\begin{equation}
K_{ijk\ell}=\frac{1}{2V}\frac{\partial^2D_{ik}}{\partial q_j\partial q_{\ell}}\Bigg|_{\textbf{q}=\mathbf{0}}.
\end{equation}
Using the form of $K_{ijk\ell}$ for an isotropic elastic medium, we find
\begin{equation}
\begin{split}
\textbf{D}(q)&=V\left(\lambda+2\mu\right)q^2\;\hat{q}_i\hat{q}_j+V\mu q^2\left(\delta_{ij}-\hat{q}_i\hat{q}_j\right), \\
\textbf{G}(q,\omega)&=\frac{1}{V\left(\lambda+2\mu\right)q^2-w'}\;\hat{q}_i\hat{q}_j+\\
&+\frac{1}{V\mu q^2-w'}\left(\delta_{ij}-\hat{q}_i\hat{q}_j\right).
\end{split}
\end{equation}
The frequency variable is taken to be $w'\equiv m\omega^2$ for undamped dynamics and $w'\equiv i\Gamma\omega$ for overdamped dynamics (in principle one can have a bulk viscosity $\zeta$ and shear viscosity $\eta$; this eventually gets reabsorbed into a definition of the scaling variable). This decomposition is convenient because the longitudinal and transverse parts are orthogonal:
\begin{equation}
\begin{split}
\left(\hat{q}_i\hat{q}_j\right)\left(\hat{q}_j\hat{q}_k\right)&=\hat{q}_i\hat{q}_k,\\
\left(\delta_{ij}-\hat{q}_i\hat{q}_j\right)\left(\delta_{jk}-\hat{q}_j\hat{q}_k\right)&=\left(\delta_{ik}-\hat{q}_i\hat{q}_k\right),\\
\left(\hat{q}_i\hat{q}_j\right)\left(\delta_{jk}-\hat{q}_j\hat{q}_k\right)&=0.
\end{split}
\end{equation}
We can compute the integrand that appears in the CPA self-consistent equation (Equation~\eqref{eq:SC}) for this isotropic sheet:
\begin{equation}
\mathbf{D}\mathbf{G}=\frac{\left(\lambda+2\mu\right)q^2}{\left(\lambda+2\mu\right)q^2-w}\hat{q}_i\hat{q}_k+\frac{\mu q^2}{\mu q^2-w}\left(\delta_{ik}-\hat{q}_i\hat{q}_k\right),
\end{equation}
so
\begin{equation}
\textrm{Tr}\left(\mathbf{D}\mathbf{G}\right)=\frac{\left(\lambda+2\mu\right)q^2}{\left(\lambda+2\mu\right)q^2-w}+(d-1)\frac{\mu q^2}{\mu q^2-w}.
\end{equation}
where we have redefined $w\equiv w'/V=\rho\omega^2$ or $i\gamma\omega$. Using the fact that the shear modulus is being considered as the only independent modulus that is being depleted, we can write $\lambda/\lambda_{\textsc{F}}=\mu/\mu_{\textsc{F}}$ and so
\begin{equation}
\textrm{Tr}\left(\mathbf{D}\mathbf{G}\right)=\frac{\left(\lambda_{\textsc{F}}/\mu_{\textsc{F}}+2\right)\mu q^2}{\left(\lambda_{\textsc{F}}/\mu_{\textsc{F}}+2\right)\mu q^2-w}+(d-1)\frac{\mu q^2}{\mu q^2-w}.
\end{equation}
Now we rearrange the integral a bit.
\begin{equation}
\begin{split}
\frac{1}{\widetilde{z}}\fint_{\textrm{BZ}}\textrm{d}^dq\;\textrm{Tr}\left(\textbf{D}\textbf{G}\right)&=\frac{1}{\widetilde{z}}\frac{1}{s_{\textsc{BZ}}}\int_{\textsc{BZ}}\textrm{d}^dq\;\textrm{Tr}\left(\textbf{D}\textbf{G}\right)\\
&=\frac{1}{\widetilde{z}}\frac{S_{d-1}}{V_dq_{\textsc{D}}^d}\int_0^{q_{\textsc{D}}}\textrm{d}q\;q^{d-1}\;\textrm{Tr}\left(\textbf{D}\textbf{G}\right)\\
&=\frac{1}{\widetilde{z}}\frac{d}{q_{\textsc{D}}^d}\int_0^{q_{\textsc{D}}}\textrm{d}q\;q^{d-1}\;\textrm{Tr}\left(\textbf{D}\textbf{G}\right)
\end{split}
\end{equation}
where $S_{d-1}$ is the surface area of the unit $d-1$ sphere embedded in $d$-dimensional space and $V_d$ is the volume of the $d$-dimensional ball; we have used the fact that $S_{d-1}/V_d=d$ in all dimensions $d$. Now we subtract $p_c=d/\widetilde{z}$ from either side of the self-consistent equation. On the side involving just the modulus, the result is
\begin{equation}\label{eq:MDefinition}
\begin{split}
\frac{\left(p-d/\widetilde{z}\right)-\left(1-d/\widetilde{z}\right)\mu/\mu_{\textsc{F}}}{1-\mu/\mu_{\textsc{F}}}&=\frac{\delta p - \left|\delta p\right|M}{1-\left|\delta p\right|M/(1-d/\widetilde{z})}\\
&\approx \delta p - \left|\delta p\right|M,\\
M&\equiv\frac{\mu/\mu_0}{\left|\delta p\right|},\\
\mu_0&\equiv\frac{\mu_{\textsc{F}}}{\left(1-d/\widetilde{z}\right)}.
\end{split}
\end{equation}
We are justified in ignoring the denominator because it contributes terms with one higher power in $\left|\delta p\right|$ when written in terms of the scaling variables. When we eventually carefully consider the side involving the frequency, we will find terms that scale as $\left|\delta p\right|$ and $\left|\delta p\right|^{d/2}$, indicating we are safe to ignore the contribution from the denominator until $d=4$. Even above $d=4$, there are more relevant terms that dictate the appropriate scaling behavior of the real part of the modulus, and the low-frequency imaginary part cannot be fixed by including further polynomial terms like these in the self-consistent equation. We describe this physically important range of frequencies with a dangerously irrelevant variable. Note that the exponents on the invariant scaling combination combining $\mu$ and $\delta p$ are independent of dimension; this is reflected in the RG flow equations that we write down.

Now we subtract $p_c$ from the side involving the frequency in a democratic way. We write
\begin{equation}
\begin{split}
\frac{d}{\widetilde{z}}=\frac{1}{\widetilde{z}}\fint_{\textrm{BZ}}\textrm{d}^dq\;\frac{\left(\lambda+2\mu\right)q^2-w}{\left(\lambda+2\mu\right)q^2-w}+\\
+\frac{(d-1)}{\widetilde{z}}\fint_{\textrm{BZ}}\textrm{d}^dq\;\frac{\mu q^2-w}{\mu q^2-w}
\end{split}
\end{equation}
and subtract the first piece from the longitudinal contribution and the second piece from the transverse contributions. This only serves to replace the numerators $(\lambda+2\mu)q^2\rightarrow w$ and $\mu q^2\rightarrow w$, giving us
\begin{equation}
\begin{split}
\frac{1}{\widetilde{z}}\frac{d}{q_{\textsc{D}}^d}\int_0^{q_{\textsc{D}}}\textrm{d}q\;\frac{wq^{d-1}}{\left(\lambda_{\textsc{F}}/\mu_{\textsc{F}}+2\right)\mu q^2-w}+\\
+\frac{1}{\widetilde{z}}\frac{d(d-1)}{q_{\textsc{D}}^d}\int_0^{q_{\textsc{D}}}\textrm{d}q\;\frac{wq^{d-1}}{\mu q^2-w}
\end{split}
\end{equation}
or
\begin{equation}
\begin{split}
\frac{1}{\widetilde{z}}\frac{d}{q_{\textsc{D}}^d}\int_0^{q_{\textsc{D}}}\textrm{d}q\;\frac{q^{d-1}}{\frac{\left(\lambda_{\textsc{F}}/\mu_{\textsc{F}}+2\right)}{w}\mu q^2-1}+\\
+\frac{1}{\widetilde{z}}\frac{d(d-1)}{q_{\textsc{D}}^d}\int_0^{q_{\textsc{D}}}\textrm{d}q\;\frac{q^{d-1}}{\frac{\mu}{w} q^2-1}.
\end{split}
\end{equation}
We now perform a substitution $\xi=(q/q_{\textsc{D}})^2$. This serves the dual purpose of nondimensionalizing the integration variable and casting the integrals into a canonical form to be identified with a special function. The integrals become
\begin{equation}
\begin{split}
-\frac{d}{2\widetilde{z}}\int_0^{1}\textrm{d}\xi\;\frac{\xi^{d/2-1}}{1-\xi\frac{\mu}{w_{\textsc{L}}}}-\frac{d\left(d-1\right)}{2\widetilde{z}}\int_0^{1}\textrm{d}\xi\;\frac{\xi^{d/2-1}}{1-\xi\frac{\mu}{w_{\textsc{T}}}},\\
w_{\textsc{L}}\equiv\frac{w}{\left(\lambda_{\textsc{F}}/\mu_{\textsc{F}}+2\right)q_{\textsc{D}}^2},\;\;\;w_{\textsc{T}}\equiv\frac{w}{q_{\textsc{D}}^2}.
\end{split}
\end{equation}
Now each of these integrals are of the form
\begin{equation}
\begin{split}
\int_0^1\textrm{d}\xi\frac{\xi^{d/2-1}}{1-\xi z}&= \\
=\int_0^1\textrm{d}\xi\,&\xi^{d/2-1}\left(1-\xi\right)^{\left(d/2+1\right)-d/2-1}\left(1-\xi z\right)^{-1}\\
&=\textrm{B}\left(\frac{d}{2},1\right)\,{}_2\textrm{F}_1\left(1,\frac{d}{2};\frac{d}{2}+1;z\right)\\
&=\frac{2}{d}\,{}_2\textrm{F}_1\left(1,\frac{d}{2};\frac{d}{2}+1;z\right)
\end{split}
\end{equation}
where $\textrm{B}(z_1,z_2)$ is the beta function and ${}_2\textrm{F}_1(a,b;c;z)$ is the ordinary hypergeometric function, as can be verified in $9.111$ of \cite{GRIntegrals} (and using the beta function identity $\textrm{B}\left(z,1\right)=1/z$). The frequency-dependent part of Equation~\eqref{eq:SC} in the isotropic case (with $p_c$ subtracted out) is then exactly
\begin{equation}
\begin{split}
-\frac{1}{\widetilde{z}}\,{}_2\textrm{F}_1\left(1,\frac{d}{2};\frac{d}{2}+1;\frac{\mu}{w_{\textsc{L}}}\right)-\\
-\frac{(d-1)}{\widetilde{z}}\,{}_2\textrm{F}_1\left(1,\frac{d}{2};\frac{d}{2}+1;\frac{\mu}{w_{\textsc{T}}}\right).
\end{split}
\end{equation}
The parameter $\mu$ is found self-consistently and is some complex number in the appropriate scaling limit. The imaginary parts of the viscoelastic moduli are nonpositive to respect the causality of the Green's function. As mentioned in the main text, taking the scaling limit amounts to sending the argument of the hypergeometric functions to $\infty$, so we are interested in expansions of ${}_2\textrm{F}_1\left(\alpha,\beta;\gamma;z\right)$ about its branch point at $z=\infty$. For this we look at the second identity in $9.132$ of \cite{GRIntegrals}, assume $d$ is not even for now and write 
\begin{equation}
\begin{split}
\frac{\Gamma\left(\frac{d}{2}\right)^2}{\Gamma\left(\frac{d}{2}+1\right)\Gamma\left(\frac{d}{2}-1\right)}{}_2\textrm{F}_1\left(1,\frac{d}{2};\frac{d}{2}+1;\frac{\mu}{w_{\textsc{L}}}\right)=\\
\left(-\frac{w_{\textsc{L}}}{\mu}\right){}_2\textrm{F}_1\left(1,1-\frac{d}{2};2-\frac{d}{2};\frac{w_{\textsc{L}}}{\mu}\right)+\\
+\Gamma\left(\frac{d}{2}\right)^2\frac{\Gamma\left(1-\frac{d}{2}\right)}{\Gamma\left(\frac{d}{2}-1\right)}\left(-\frac{w_{\textsc{L}}}{\mu}\right)^{d/2}.
\end{split}
\end{equation}
The hypergeometric function involving $z=w_{\textsc{L}}/\mu$ is $1$ for $z=0$ and can otherwise be expanded in a power series in $z$, which contributes terms higher-order in $\left|\delta p\right|$. Writing out the side with the frequency dependence now, grouping terms with the same powers of $w_{\textsc{L}/\textsc{T}}$:
\begin{equation}
\begin{split}
\approx\frac{1}{\widetilde{z}}\frac{\Gamma\left(\frac{d}{2}+1\right)\Gamma\left(\frac{d}{2}-1\right)}{\Gamma\left(\frac{d}{2}\right)^2}\frac{w_{\textsc{L}}+(d-1)w_{\textsc{T}}}{\mu}-\\
-\frac{1}{\widetilde{z}}\Gamma\left(\frac{d}{2}+1\right)\Gamma\left(1-\frac{d}{2}\right)\frac{w_{\textsc{L}}^{d/2}+(d-1)w_{\textsc{T}}^{d/2}}{(-\mu)^{d/2}}.
\end{split}
\end{equation}
The scaling variables asymptotically close to the critical point are now ready to be defined, but they depend upon whether we are above or below $d=2$. We investigate each case separately below.

\subsection{Scaling variables above 2 dimensions}
First, assume $d>2$ so that the first term contributes the leading-order frequency behavior. Then we define a scaling for the frequency $f$:
\begin{equation}
\left|\delta p\right|F\equiv\frac{1}{\widetilde{z}}\frac{\Gamma\left(\frac{d}{2}+1\right)\Gamma\left(\frac{d}{2}-1\right)}{\Gamma\left(\frac{d}{2}\right)^2}\frac{w_{\textsc{L}}+(d-1)w_{\textsc{T}}}{\mu}.
\end{equation}
This corresponds to:
\begin{equation}\label{eq:FDefinition}
\begin{split}
F&=\frac{f/f_0}{\left|\delta p\right|},\\
f_0&=q_{\textsc{D}}^2\frac{\Gamma\left(\frac{d}{2}\right)^2}{\Gamma\left(\frac{d}{2}+1\right)\Gamma\left(\frac{d}{2}-1\right)}\frac{\widetilde{z}/\mu_{\textsc{F}}}{\frac{1}{\lambda_{\textsc{F}}+2\mu_{\textsc{F}}}+\left(d-1\right)\frac{1}{\mu_{\textsc{F}}}},
\end{split}
\end{equation}
essentially setting $f_0=c^2q_{\textsc{D}}^2$, where $c$ is a weighted combination of the longitudinal and transverse sound speeds in the undepleted membrane.

This definition of the scaling variable also makes the first term of the frequency-dependent side of the self-consistent equation $\left|\delta p\right|F$. These scaling variables are then inserted into the other term, and the variable $G$ is defined so that the final term is $+\left|\delta p\right|G(-F)^{d/2}$ (note that $\Gamma(1-d/2)$ is negative for $2<d<4$). This gives us
\begin{equation}\label{eq:GDefinition}
\begin{split}
G&\equiv g/g_0\left|\delta p\right|^{d/2-1},\\
g/g_0&=\widetilde{z}^{d/2-1}\frac{\Gamma\left(\frac{d}{2}+1\right)\left(-\Gamma\left(1-\frac{d}{2}\right)\right)\Gamma\left(\frac{d}{2}\right)^d}{\left(\Gamma\left(\frac{d}{2}+1\right)\Gamma\left(\frac{d}{2}-1\right)\right)^{d/2}}\times \\
&\times\left(\frac{\left(\frac{1}{\lambda_{\textsc{F}}+2\mu_{\textsc{F}}}\right)^{d/2}+\left(d-1\right)\left(\frac{1}{\mu_{\textsc{F}}}\right)^{d/2}}{\left(\left(\frac{1}{\lambda_{\textsc{F}}+2\mu_{\textsc{F}}}\right)+\left(d-1\right)\left(\frac{1}{\mu_{\textsc{F}}}\right)\right)^{d/2}}\right).
\end{split}
\end{equation}
The large prefactor involving $\Gamma$ functions can be simplified to bring the expression into the form
\begin{equation}
\begin{split}
g/g_0&=-\widetilde{z}^{d/2-1}\left(\frac{d-2}{d}\right)^{d/2}\frac{\pi d}{2}\csc\left(\frac{\pi d}{2}\right)\times \\
&\times\left(\frac{\left(\frac{1}{\lambda_{\textsc{F}}+2\mu_{\textsc{F}}}\right)^{d/2}+\left(d-1\right)\left(\frac{1}{\mu_{\textsc{F}}}\right)^{d/2}}{\left(\left(\frac{1}{\lambda_{\textsc{F}}+2\mu_{\textsc{F}}}\right)+\left(d-1\right)\left(\frac{1}{\mu_{\textsc{F}}}\right)\right)^{d/2}}\right)
\end{split}
\end{equation}
where the term involving $\csc$ is negative for $2<d<4$. The self-consistent equation for $d>2$ defining the universal scaling function is then (dividing both sides by $\left|\delta p\right|$)
\begin{equation}
\pm1-M=F+G\left(-F\right)^{d/2},
\end{equation}
as claimed in the main text.

\subsection{2 dimensions as a limit}

To take the limit $d\rightarrow 2$, we first factor out one power of $-F$ from the self-consistent equation valid for $d>2$. We then define
\begin{equation}
F'\equiv\frac{F}{\Gamma\left(\frac{d}{2}-1\right)}
\end{equation}
to remove the divergence in the definition of $F$ while retaining the same critical exponents (for now). The equation becomes
\begin{equation}
\begin{split}
\pm1-M&=\\-F'&\left(\left(-F'\Gamma\left(\frac{d}{2}-1\right)\right)^{d/2-1}G-1\right)\Gamma\left(\frac{d}{2}-1\right).
\end{split}
\end{equation}
Bring $G$ inside the power of $d/2-1$ and write it as $G^{2/(d-2)}=\left(g/g_0\right)^{2/(d-2)}\left|\delta p\right|$. Then define $(g/g_0)'^{2/(d-2)}\equiv \left(g/g_0\right)^{2/(d-2)}\Gamma\left(d/2-1\right).$ The self-consistent equation becomes
\begin{equation}
\begin{split}
\pm1-M&=\\-F'&\left(\left(-(g/g_0)'^{\frac{2}{2-d}}F'\left|\delta p\right|\right)^{d/2-1}-1\right)\Gamma\left(\frac{d}{2}-1\right),
\end{split}
\end{equation}
as claimed in the main text. Now we need only to compute $g_2/g_{02}=\lim_{d\rightarrow2^+}(g/g_0)'^{2/(d-2)}$. This yields
\begin{equation}
g_2/g_{02}=\widetilde{z}\,\frac{\mu_{\textsc{F}}^{\frac{\mu_{\textsc{F}}}{\lambda_{\textsc{F}}+3\mu_{\textsc{F}}}}\left(\lambda_{\textsc{F}}+2\mu_{\textsc{F}}\right)^{\frac{\lambda_{\textsc{F}}+2\mu_{\textsc{F}}}{\lambda_{\textsc{F}}+3\mu_{\textsc{F}}}}}{\lambda_{\textsc{F}}+3\mu_{\textsc{F}}}.
\end{equation}
The self-consistent equation in $d=2$ is then
\begin{equation}
\pm1-M=-F'\log\left(-(g_2/g_{02})F'\left|\delta p\right|\right),
\end{equation}
but as mentioned in the main text, this is not written in terms of the proper scaling variables. This result can also be derived without hypergeometric functions by directly performing the integral in $d=2$, but it becomes less clear how  this result is continuously connected to $d=3$, or $d<2$. It can also be understood (equivalently) as the branch point of the hypergeometric function at $z=\infty$ transforming from a power-law like branch to a logarithmic branch for $d$ even. This kind of resonance behavior, where the term with the exponent $d/2$ interacts with another term with an integer power to give a logarithm, happens in every even dimension. For even dimensions above $2$, the logarithm is associated with a correction that vanishes close to the critical point, but it is necessary to retain to understand the low-frequency imaginary part of the viscoelastic modulus.

In the right scaling variables, we can absorb this additional logarithmic divergence into the definition of the invariant scaling combination involving $f$ in $d=2$, giving
\begin{equation}
\pm1-M=F_2,
\end{equation}
by analogy to the scaling variable for $f$ in all dimensions $d>2$. We can examine what this implies about how $f$ scales with $\delta p$. We have 
\begin{equation}
F_2=-F'\log\left(-(g_2/g_{02}) F'\left|\delta p\right|\right)
\end{equation}
or
\begin{equation}
(g_2/g_{02})F_2\left|\delta p\right|=-(g_2/g_{02})F'\left|\delta p\right|\log\left(-(g_2/g_{02}) F'\left|\delta p\right|\right)
\end{equation}
where $F_2$ is the invariant scaling combination. We invert this with the $W$ function:
\begin{equation}
-(g_2/g_{02})F'\left|\delta p\right|=\frac{(g_2/g_{02})F_2\left|\delta p\right|}{W\left((g_2/g_{02})F_2\left|\delta p\right|\right)}
\end{equation}
reinstalling the definition of $F'=f/f_{0}'/\left|\delta p\right|$, we have (up to constants)
\begin{equation}
f\sim\frac{(g_2/g_{02})F_2\left|\delta p\right|}{W\left((g_2/g_{02})F_2 \left|\delta p\right|\right)}
\end{equation}
in $d=2$, which is also supported by the RG flow equations.
\subsection{Scaling variables below 2 dimensions}
Below $2$ dimensions, the term involving $w_{\textsc{L}/\textsc{T}}^{d/2}$ is dominant at low frequencies. We define a scaling variable for the frequency $f$:
\begin{equation}
\begin{split}
\left|\delta p\right|\left(-F_d\right)^{d/2}\equiv \frac{1}{\widetilde{z}}\Gamma\left(\frac{d}{2}+1\right)\Gamma\left(1-\frac{d}{2}\right)\times\\
\times\frac{w_{\textsc{L}}^{d/2}+\left(d-1\right)w_{\textsc{T}}^{d/2}}{\left(-\mu\right)^{d/2}}
\end{split}
\end{equation}
Note that for $d<2$, $\Gamma\left(1-d/2\right)$ is now positive. This corresponds to:
\begin{equation}\label{eq:FDefinitionLower}
\begin{split}
F_d&=\frac{f/f_{0d}}{\left|\delta p\right|^{2/d}},\\
f_{0d}&= q_{\textsc{D}}^2\frac{1}{\Gamma\left(\frac{d}{2}+1\right)^{2/d}\Gamma\left(1-\frac{d}{2}\right)^{2/d}}\times\\
\times&\frac{\widetilde{z}^{2/d}/\mu_{\textsc{F}}}{\left(\left(\frac{1}{\lambda_{\textsc{F}}+2\mu_{\textsc{F}}}\right)^{d/2}+\left(d-1\right)\left(\frac{1}{\mu_{\textsc{F}}}\right)^{d/2}\right)^{2/d}},
\end{split}
\end{equation}
another constant that is similar to $c^2q_{\textsc{D}}^2$. This sets the first term in the self-consistent equation to be $-\left|\delta p\right|\left(-F_d\right)^{d/2}$. We again insert the definitions of the scaling variable $F_d$ into the remaining term, and define an additional variable $G_d$ so that the final term is $-\left|\delta p\right|G_dF_d$. This gives us
\begin{equation}\label{eq:GDefinitionLower}
\begin{split}
G_d&\equiv (g_d/g_{0d})\left|\delta p\right|^{2/d-1},\\
g_d/g_{0d}&=\widetilde{z}^{2/d-1}\frac{-\Gamma\left(\frac{d}{2}-1\right)}{\Gamma\left(\frac{d}{2}\right)^2\Gamma\left(\frac{d}{2}+1\right)^{2/d-1}\Gamma\left(1-\frac{d}{2}\right)^{2/d}}\times\\
&\times\frac{\left(\frac{1}{\lambda_{\textsc{F}}+2\mu_{\textsc{F}}}\right)+\left(d-1\right)\left(\frac{1}{\mu_{\textsc{F}}}\right)}{\left(\left(\frac{1}{\lambda_{\textsc{F}}+2\mu_{\textsc{F}}}\right)^{d/2}+\left(d-1\right)\left(\frac{1}{\mu_{\textsc{F}}}\right)^{d/2}\right)^{2/d}}.
\end{split}
\end{equation}
Note that for $d<2$, $\Gamma\left(d/2-1\right)$ is now negative. The self-consistent equation for $d<2$ defining the universal scaling function is then (dividing both sides by $\left|\delta p\right|$)
\begin{equation}
\pm1-M=-\left(-F_d\right)^{d/2}-G_dF_d,
\end{equation}
as claimed in the main text.
\subsection{Scaling in even dimensions greater than 2}
To understand the scaling  in even dimensions $d>2$ one needs to retain three terms from the evaluation of the integral: the dominant piece from the hypergeometric function, the non-analytic piece going as $\sim\left(-w/\mu\right)^{d/2}$, and the $\lfloor d/2\rfloor$th term of the power series expansion of the hypergeometric function (note that the first and third of these coincide when $d=2$). The latter two terms will interact in the limit $d/2\rightarrow k^+$ for $k$ an integer $>1$ and produce a logarithmic singularity in a way similar to the $2$ dimensional case. Keeping these terms, and writing things in terms of scaling variables valid above $d>2$ gives
\begin{equation}
\begin{split}
\pm1-M&=F-\widehat{G}_d\left(-F\right)^{\lfloor d/2\rfloor}+G\left(-F\right)^{d/2}.
\end{split}
\end{equation}
Here $\widehat{G}_d$ is an invariant scaling combination associated with some other irrelevant variable $\sim\left|\delta p\right|^{\lfloor d/2\rfloor-1}$. As in dimension $2$, pull out a factor of $\widehat{G}_d\left(-F\right)^{\lfloor d/2\rfloor}$ from the last two terms, and one finds
\begin{equation}
\begin{split}
\pm1-M&=F+\\
+\widehat{G}_d\left(-F\right)^{\lfloor d/2\rfloor}&\left(\frac{G}{\widehat{G}_d}\left(-F\right)^{d/2-\lfloor d/2\rfloor}-1\right).
\end{split}
\end{equation}
Similar to the case in $d=2$, we can redefine $\widehat{G}_d$ to $\widehat{G}_d'$ to pull out a divergence in even dimensions, and take the limit as $d/2\rightarrow\lfloor d/2\rfloor$ to recover (making the $\delta p$ dependence of the irrelevant variables explicit)
\begin{equation}
\begin{split}
\pm1-M&=F+\\
+c_d\left|\delta p\right|^{\lfloor d/2\rfloor-1}\left(-F\right)^{\lfloor d/2\rfloor}&\log\left(-g_dF\left|\delta p\right|\right).
\end{split}
\end{equation}
This shows that the scaling exponents for all relevant variables in $d>2$ are the same; the $F$ term sets the scaling and the remaining term is a correction that vanishes. The only novelty is that the non-analyticity that fixes the low-frequency imaginary part is now a logarithmic singularity, rather than a power-law singularity. As in other $d>2$ this correction term must be retained to capture the low-frequency dissipation in the microscopically undamped case. As in other integer dimensions, the self-consistent integral can be expressed directly in terms of rational functions and logarithms (without referring to special functions) to verify these formulas.
\subsection{Density of states scaling}
\begin{figure}[ht]
\begin{center}
\includegraphics[width=\linewidth]
{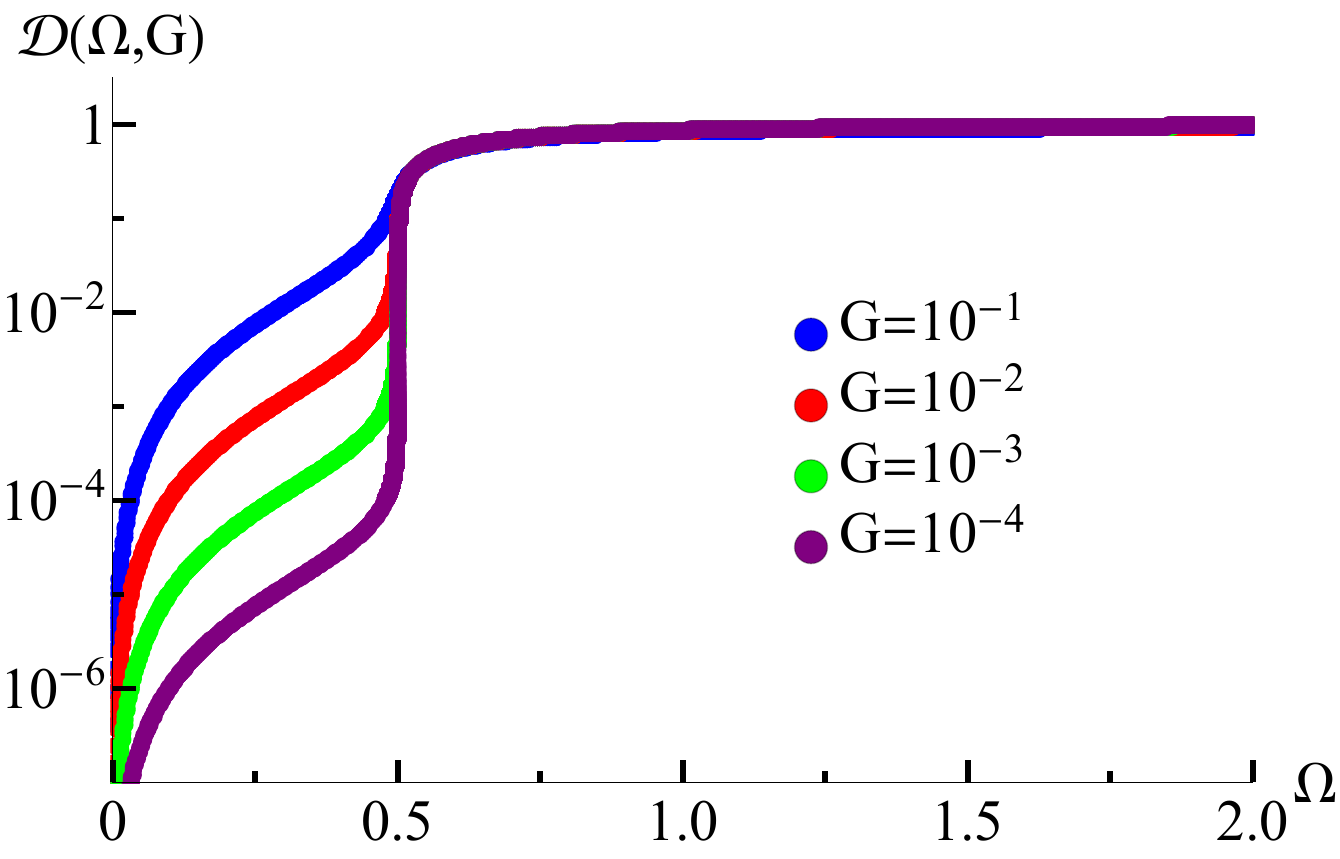}
\end{center}
\caption{\textbf{Density of states for a nearly floppy 3D viscoelastic membrane}. The dangerously irrelevant variable controls the low-frequency density of states in the undamped case. The rescaled density of states has a square-root cusp at $G=0$ $(\delta p=0)$~\cite{LiarteXOL19,LiarteTSCCS22}; as discussed in section~\ref{subsec:3d}, the continuum CPA, and our universal scaling limit, do not show the $\omega^4$ contribution to the density of states found from quasilocalized modes~\cite{Quasilocalized2018}. See Figure~\ref{BosonPeak} for the excess density of states $\mathcal{D}\left(\Omega,G\right)/\Omega^2$.}
\label{DOS3DG}
\end{figure}

The density of states is given in the undamped case as
\begin{equation}
D\left(\omega\right)=\frac{\omega}{\pi}\int_{\textrm{BZ}}\textrm{d}^dq\,\textrm{Im}\left(\textrm{Tr}\left(\textbf{G}\right)\right).
\end{equation}
This is evaluated for an isotropic system similarly to the previous section:
\begin{equation}
\begin{split}
D\left(\omega\right)=\frac{\omega}{\pi V}\textrm{Im}\left[\int_{0}^{q_{\textsc{D}}}\textrm{d}q\,\frac{S_{d-1}q^{d-1}}{\left(\lambda_{\textsc{F}}/\mu_{\textsc{F}}+2\right)\mu q^2-w}+\right.\\
\left.+\left(d-1\right)\int_{0}^{q_{\textsc{D}}}\textrm{d}q\,\frac{S_{d-1}q^{d-1}}{\mu q^2-w}\right].
\end{split}
\end{equation}
A trick to eliminate more tedious manipulation is to multiply and divide by a particular factor:
\begin{equation}
\begin{split}
D\left(\omega\right)=\frac{\omega}{\pi }\frac{\widetilde{z}V_dq_{\textsc{D}}^d}{Vw}\textrm{Im}\left[\frac{d}{\widetilde{z}q_{\textsc{D}}^d}\int_{0}^{q_{\textsc{D}}}\textrm{d}q\,\frac{wq^{d-1}}{\left(\lambda_{\textsc{F}}/\mu_{\textsc{F}}+2\right)\mu q^2-w}+\right.\\
\left.+\left(d-1\right)\frac{d}{\widetilde{z}q_{\textsc{D}}^d}\int_{0}^{q_{\textsc{D}}}\textrm{d}q\,\frac{wq^{d-1}}{\mu q^2-w}\right].
\end{split}
\end{equation}
The term in brackets can be identified, to leading order in the scaling variables, as
\begin{equation}
D\left(\omega\right)\approx\frac{\omega}{\pi}\frac{\widetilde{z}V_dq_{\textsc{D}}^d}{Vw}\textrm{Im}\left[\delta p-\left|\delta p\right|M\right]=-\left|\delta p\right|\frac{\widetilde{z}V_dq_{\textsc{D}}^d}{\pi m\omega}\textrm{Im}\left[M\right]
\end{equation}
For $d>2$, this gives the following scaling form (written in terms of scaling variables for $\omega$ rather than for $f=w/\mu$):
\begin{equation}
\begin{split}
\mathcal{D}\left(\Omega,G\right)&=-\frac{\textrm{Im}\left[M\left(\Omega,G\right)\right]}{\Omega},\\
\mathcal{D}&\equiv\frac{D}{D_0}.
\end{split}
\end{equation}
This remarkably simple scaling form is true within the CPA~\cite{DeGiuli2014}, but not in general. Setting $G=0$ allows us to exactly evaluate this; in this scaling limit there is only a nonzero density of states for $\Omega>1/2$ and it is flat at high frequency. We have
\begin{equation}
\mathcal{D}\left(\Omega,0\right)=\frac{\sqrt{4\Omega^2-1}}{2\Omega}.
\end{equation}
The density of states is shown for $d=3$ and various values of $G$ in Figure~\ref{DOS3DG}.

\begin{figure}[ht]
\begin{center}
\includegraphics[width=\linewidth]
{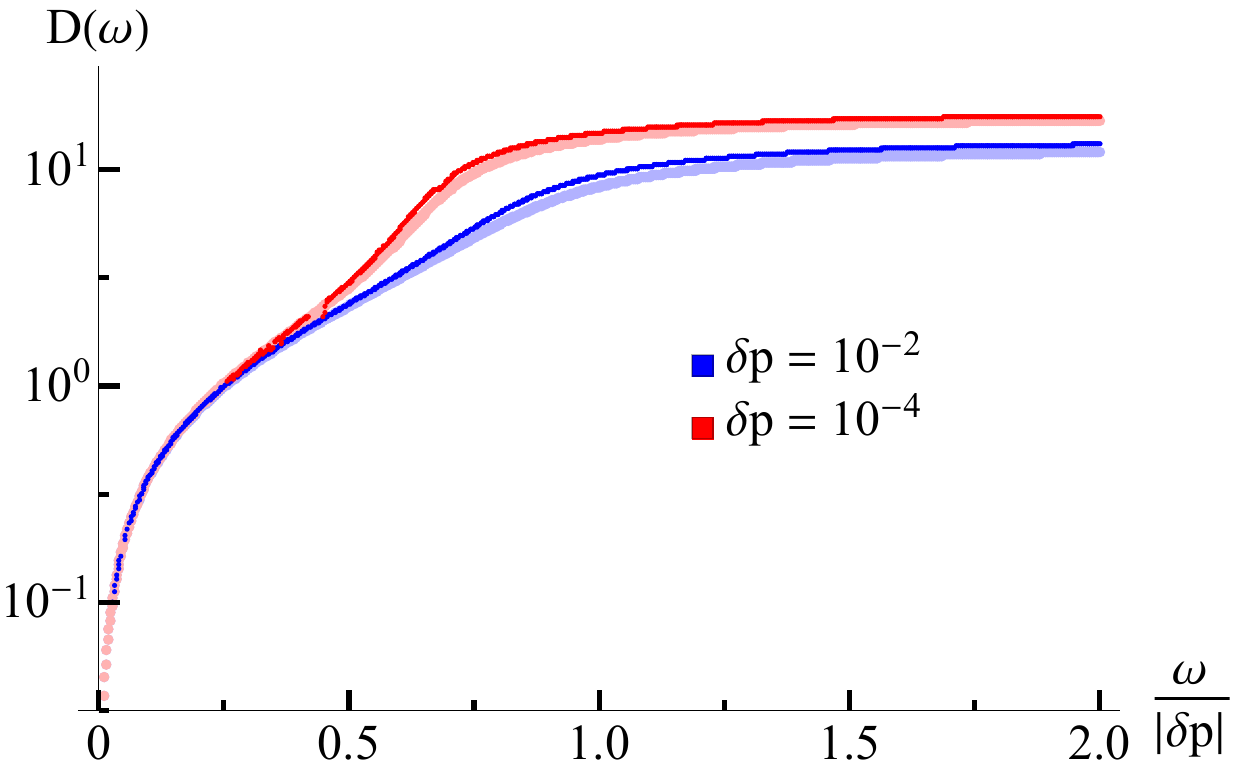}
\end{center}
\caption{\textbf{Density of states for the diluted triangular lattice computed within the CPA}. We plot rescaled numerical solutions (thin lines) against our scaling solutions (thick lines) at two distances from the critical point $\delta p=\left\{10^{-2},10^{-4}\right\}$, again demonstrating nice agreement with the scaling form. The shift in the rescaled frequency $\Omega_{\textsc{T}}^*$ where we cross over to a flat density of states and the vertical shift in the location of the plateau are both due to logarithmic corrections present in the upper critical dimension.}
\label{DOS2DComp}
\end{figure}

For $d<2$, we have:
\begin{equation}
\begin{split}
\mathcal{D}_d\left(\Omega_d,G_d\right)&=-\frac{\textrm{Im}\left[M\left(\Omega_d,G_d\right)\right]}{\Omega_d},\\
\mathcal{D}_d&\equiv\frac{D}{D_{0d}}\left|\delta p\right|^{1/d-1/2}.
\end{split}
\end{equation}
It can also be written for $d>2$ as
\begin{equation}\label{eq:DDefinition}
\begin{split}
\mathcal{D}\left(F,G\right)&=-\frac{\textrm{Im}\left[M\left(F,G\right)\right]}{M^{1/2}F^{1/2}},\\
\mathcal{D}&\equiv\frac{D}{D_0'}
\end{split}
\end{equation}
and for $d<2$ as
\begin{equation}\label{eq:DDefinitionLower}
\begin{split}
\mathcal{D}_d\left(F_d,G_d\right)&=-\frac{\textrm{Im}\left[M\left(F_d,G_d\right)\right]}{M^{1/2}F_d^{1/2}},\\
\mathcal{D}_d&\equiv\frac{D}{D_{0d}'}\left|\delta p\right|^{1/d-1/2}.
\end{split}
\end{equation}
In $d=2$, we again have logarithmic corrections:
\begin{equation}
\begin{split}
\mathcal{D}_2\left(F_2\right)&=-\frac{\textrm{Im}\left[M\left(F_2\right)\right]}{M^{1/2}F_2^{1/2}},\\
\mathcal{D}_2&\equiv\frac{D/D_{02}}{\sqrt{W\left(x(g)\left|\delta p\right|\right)}}.
\end{split}
\end{equation}
For the triangular lattice in two dimensions (see next section for all specifics of lattice constants and microscopic stiffness parameters), we can reinstall the non-universal constants in the definitions of the scaling variables to make a direct comparison to the numerically determined density of states. To highlight the logarithmic shifts, we define $M_{\textsc{T}}\equiv\mu/\left|\delta p\right|$ and $\Omega_{\textsc{T}}\equiv\omega/\left|\delta p\right|$. In these variables, the asymptotic form of the density of states for an elastic sheet with the same long-wavelength parameters as the triangular lattice is
\begin{equation}
D_2=-\frac{32\pi}{3\sqrt{3}}\frac{\textrm{Im}\left(M_{\textsc{T}}\right)}{\Omega_{\textsc{T}}},
\end{equation}
where $M_{\textsc{T}}$ is the solution to
\begin{equation}
\pm1-\frac{4}{3\sqrt{3}}M_{\textsc{T}}=-\frac{1}{3\pi\sqrt{3}}\frac{\Omega_{\textsc{T}}^2}{M_{\textsc{T}}}\log\left(-\frac{3^{1/4}}{4\pi}\frac{\Omega_{\textsc{T}}^2}{M_{\textsc{T}}}\left|\delta p\right|\right).
\end{equation}
(See Appendix~\ref{AppD} for details.) The plot of the comparison is shown in Figure~\ref{DOS2DComp} for $\delta p=10^{-2}$ and $\delta p=10^{-4}$. The slow leftward drift of the onset of the plateau in the DOS and the upward drift of the location of the plateau in the DOS are both related to $d=2$ being the upper critical dimension of the theory.

One often identifies the excess soft modes in glassy systems by plotting the density of vibrational states divided by the expected form from a Debye model of a crystal. This amounts to dividing the universal scaling function $\mathcal{D}$ by $\Omega^{d-1}$ in dimension $d$; as the density of states is flat for $\Omega\gtrsim1$, these plots unsurprisingly show a bump (Figure~\ref{BosonPeak}).

\subsection{Green's function scaling}
For completeness, we report the scaling form of the Green's function. For the case of the continuous transition, we write the long-wavelength form associated with an elastic sheet:
\begin{equation}
\begin{split}
V\textbf{G}(q,\omega)&=\frac{1}{\left(\lambda+2\mu\right)q^2-w}\;\hat{q}_i\hat{q}_j+\\
&+\frac{1}{\mu q^2-w}\left(\delta_{ij}-\hat{q}_i\hat{q}_j\right).
\end{split}
\end{equation}
This can be rewritten in terms of $f$:
\begin{equation}
\begin{split}
V\textbf{G}(q,\omega)&=\frac{1}{\mu\left(\left(\lambda_{\textsc{F}}/\mu_{\textsc{F}}+2\right)q^2-f\right)}\;\hat{q}_i\hat{q}_j+\\
&+\frac{1}{\mu\left(q^2-f\right)}\left(\delta_{ij}-\hat{q}_i\hat{q}_j\right)
\end{split}
\end{equation}
so $q^2$ and $f$ have identical scaling asymptotics in all dimensions (this can also be seen from the RG flow Equations~\eqref{eq:RGflows}). The non-universal constants that set the scale for $q$ differ for the transverse and longitudinal parts, but the form of the universal scaling function is identical: in dimensions $d>2$,
\begin{equation}\label{eq:GreenDefinition}
\begin{split}
\mathcal{G}\left(Q,F,G\right)&=\frac{1}{M\left(F,G\right)\left(Q^2-F\right)}\\
\mathcal{G}\equiv\frac{G_{\textsc{L}/\textsc{T}}}{G^{\textsc{L}/\textsc{T}}_0}&\left|\delta p\right|^2,\;\;\;Q\equiv\frac{q/q^{\textsc{L}/\textsc{T}}_0}{\left|\delta p\right|^{1/2}}.
\end{split}
\end{equation}
Similarly, in dimensions $d<2$, we have
\begin{equation}\label{eq:GreenDefinitionLower}
\begin{split}
\mathcal{G}_d\left(Q_d,F_d,G_d\right)&=\frac{1}{M\left(F_d,G_d\right)\left(Q_d^2-F_d\right)}\\
\mathcal{G}_d\equiv\frac{G_{\textsc{L}/\textsc{T}}}{G^{\textsc{L}/\textsc{T}}_{0d}}&\left|\delta p\right|^{1+2/d},\;\;\;Q_d\equiv\frac{q/q^{\textsc{L}/\textsc{T}}_{0d}}{\left|\delta p\right|^{1/d}}.
\end{split}
\end{equation}
Finally, in $d=2$, we have
\begin{equation}
\begin{split}
\mathcal{G}_2\left(Q_2,F_2\right)&=\frac{1}{M\left(F_2\right)\left(Q_2^2-F_2\right)}\\
\mathcal{G}_2&\equiv\frac{G_{\textsc{L}/\textsc{T}}}{G^{\textsc{L}/\textsc{T}}_{02}}\left|\delta p\right|^{2}W\left(x(g)\left|\delta p\right|\right),\\
Q_2&\equiv\frac{q/q^{\textsc{L}/\textsc{T}}_{02}}{\left|\delta p\right|^{1/2}}\sqrt{W\left(x(g)\left|\delta p\right|\right)}.
\end{split}
\end{equation}
These forms imply diverging length scales at the transition: $\ell_c\sim\left|\delta p\right|^{-1/2}$ for $d>2$, $\ell_c\sim\left|\delta p\right|^{-1/d}$ for $d<2$, and $\ell_c\sim\left|\delta p\right|^{-1/2}\left|\log\left|\delta p\right|\right|^{1/2}$ in $d=2$ (noted also in Table~\ref{tab:CritExp}). If a phonon has a wavelength shorter than $\ell_c$, it is strongly overdamped. In the case of jamming, the transverse shear mode is associated with a diverging length scale of the type analyzed here and elsewhere~\cite{LiarteTSCCS22,LiarteTSCCSnn2} but the longitudinal mode has a different scaling.

\section{Details of the triangular lattice numerics}\label{AppD}
Here we directly compare our continuum, isotropic expansion of the CPA to the lattice CPA for the bond-diluted triangular lattice. (As noted earlier, the rigidity transition of the diluted triangular lattice is not described correctly by our CPA analysis~\cite{WillWIP}. The static critical exponents for the triangular lattice have lengths which scale as $\left|\delta p\right|^{-\nu}$ and moduli which scale as $\left|\delta p\right|^f$, with $\nu \sim 1.3 \pm 0.2$ and $f \sim 2.2 \pm 0.3$; CPA predicts $\nu = 1/2$ and $f=1$, and log corrections.)

To make this comparison with no numerically determined fitting parameters, we must know the values of the \textit{nonuniversal} constants $\mu_0$, $f_{02}\equiv \lim_{d\rightarrow 2^+}f_0\Gamma\left(d/2-1\right)$, and $g_2/g_{02}$. We take the triangular lattice with nearest-neighbor bonds of strength $k$ and bond length $a=1$. The dynamical matrix is 
\begin{equation}
\textbf{D}=\begin{pmatrix}
D_{xx} & D_{xy} \\
D_{yx} & D_{yy}
\end{pmatrix}
\end{equation}
with 
\begin{equation}
\begin{split}
D_{xx}&=4k\Bigg(
\sin^2\left(\frac{q_x}{2}\right)+\frac{1}{4}\sin^2\left(\frac{q_x}{4}+\frac{\sqrt{3}}{4}q_y\right)+\\
&+\frac{1}{4}\sin^2\left(\frac{q_x}{4}-\frac{\sqrt{3}}{4}q_y\right)\Bigg),
\end{split}
\end{equation}
\begin{equation}
\begin{split}
D_{xy}=D_{yx}&=\sqrt{3}k\Bigg(\sin^2\left(\frac{q_x}{4}+\frac{\sqrt{3}}{4}q_y\right)-\\
&-\sin^2\left(\frac{q_x}{4}-\frac{\sqrt{3}}{4}q_y\right)\Bigg),
\end{split}
\end{equation}
\begin{equation}
\begin{split}
D_{yy}&=3k\Bigg(\sin^2\left(\frac{q_x}{4}+\frac{\sqrt{3}}{4}q_y\right)+\\
&+\sin^2\left(\frac{q_x}{4}-\frac{\sqrt{3}}{4}q_y\right)\Bigg).
\end{split}
\end{equation}
The Brillouin zone is a hexagon with side length $4\pi/3$, so the area of the Brillouin zone is $s_{\textsc{BZ}}=8\pi^2/\sqrt{3}$ and so $q_{\textsc{D}}=\sqrt{8\pi/\sqrt{3}}$. For small $q_x$ and $q_y$, we expand the dynamical matrix to quadratic order and find
\begin{equation}
\begin{split}
D_{xx}&=\frac{3}{8}k\left(q_x^2+q_y^2\right)+\frac{3}{4}kq_x^2+\mathcal{O}(q^4),\\
D_{xy}=D_{yx}&=\frac{3}{4}kq_xq_y+\mathcal{O}(q^4),\\
D_{yy}&=\frac{3}{8}k\left(q_x^2+q_y^2\right)+\frac{3}{4}kq_y^2+\mathcal{O}(q^4).
\end{split}
\end{equation}
The long-wavelength isotropic form of the dynamical matrix is
\begin{equation}
\textbf{D}(q)=V\left(\lambda+2\mu\right)q^2\;\hat{q}_i\hat{q}_j+V\mu q^2\left(\delta_{ij}-\hat{q}_i\hat{q}_j\right).
\end{equation}
with $V=\sqrt{3}/2$ (a hexagon with side length $1/\sqrt{3}$). Comparing the two, we find that the triangular lattice is isotropic at long wavelengths with $\lambda=\mu=\sqrt{3}k/4$. The triangular lattice has an average of $\widetilde{z}=3$ bonds per site, identifying $p_c=2/3$. This is all of the information we need to make the comparison between the triangular lattice CPA numerics and the asymptotic scaling forms for the weakened isotropic elastic sheet. Measuring the stiffnesses in units of $k$, we have \begin{equation}
\begin{split}
\lambda_{\textsc{F}}=\mu_{\textsc{F}}&=\frac{\sqrt{3}}{4},\\
\widetilde{z}&=3,\\
d&=2.
\end{split}
\end{equation}
This leads to
\begin{equation}
\begin{split}
\mu_0\equiv\frac{\mu_{\textsc{F}}}{1-d/\widetilde{z}}&=\frac{3\sqrt{3}}{4},\\
f_{02}&=\frac{6\pi\sqrt{3}}{\rho},\\
g_2/g_{02}&=\frac{3^{7/4}}{4}.
\end{split}
\end{equation}
Setting $\rho$ arbitrarily to $2$ (which sets the microscopic mass $m$ to $\sqrt{3}$), we have an \textit{ansatz} for the scaling form of the viscoelastic modulus of the diluted triangular lattice, rescaling to $M_{\textsc{T}}=\mu/\left|\delta p\right|$ and $\Omega_{\textsc{T}}=\omega/\left|\delta p\right|$:
\begin{equation}\label{eq:2DTriangularSCM}
\pm1-\frac{4}{3\sqrt{3}}M_{\textsc{T}}=-\frac{1}{3\pi\sqrt{3}}\frac{\Omega_{\textsc{T}}^2}{M_{\textsc{T}}}\log\left(-\frac{3^{1/4}}{4\pi}\frac{\Omega_{\textsc{T}}^2}{M_{\textsc{T}}}\left|\delta p\right|\right).
\end{equation}
This is compared with the full CPA for the bond-diluted triangular lattice:
\begin{equation}
\frac{p-k/k_{\textsc{F}}}{1-k/k_{\textsc{F}}}=\frac{1}{\widetilde{z}}\fint_{\textsc{BZ}}\textrm{d}^2q\;\textrm{Tr}\left(\textbf{D}\textbf{G}\right),
\end{equation}
where all expressions are for the full triangular lattice (hexagonal BZ, dynamical matrix and Green's function with triangular lattice symmetry, etc.) and $k_{\textsc{F}}=1$. The diluted triangular lattice's effective long-wavelength shear modulus is then $\mu=\sqrt{3}k/4$, and $\mu/\left|\delta p\right|$ is compared with $M_{\textsc{T}}$ (Figure~\ref{TriangularCPANumerics}). Any discrepancies that can be seen by eye are due to corrections to scaling from higher-order terms in the dynamical matrix and Green's function, which are generically anisotropic. These corrections vanish close to the critical point, and the behavior of this anisotropic triangular lattice near the critical point predicted by the CPA is well-described by this emergent isotropic theory.
\bibliographystyle{naturemag}
\normalem
\bibliography{refs,NSF,SethnaRecs}
\end{document}